\documentclass[a4paper, natbib, man, floatsintext]{apa6}

\usepackage[utf8]{inputenc}
\usepackage{graphicx}
\usepackage[disable]{todonotes}
\usepackage{authblk}
\usepackage{subfig}
\usepackage{caption}
\usepackage[hidelinks]{hyperref}
\usepackage{verbatim}
\usepackage[labelfont=bf]{caption}

\newcommand\fnote[1]{\captionsetup{font=small}\caption*{Note. #1}}

\title{\textbf{Simple and Cheap Setup for Measuring Timed Responses 
to Auditory Stimuli}}
\shorttitle{Simple and Cheap Setup for Measuring Timed Responses 
to Auditory Stimuli}

\AtBeginDocument{
    }

\leftheader{Miguel, Riera, Slezak}

\author[1,2]{Martin A. Miguel}
\author[2]{Pablo Riera}
\author[1,2]{Diego Fernandez Slezak}
\affil[1]{Universidad de Buenos Aires. Facultad de Ciencias Exactas y
Naturales. Departamento de
Computación. Buenos Aires, Argentina.}
\affil[2]{CONICET-Universidad de Buenos Aires. Instituto de Investigación en
Ciencias de la Computación
(ICC). Buenos Aires, Argentina.}

\authornote{
    We have no known conflicts of interest to disclose.

    Correspondance concerning this article should be addressed to
Martin A. Miguel. Email: mmiguel@dc.uba.ar. Address: Intendente
Güiraldes 2160 - Ciudad Universitaria - Buenos Aires, Argentina. Phone:
+541152857400
}

\abstract{
\noindent Measuring human capabilities to synchronize in time, adapt to perturbations to
timing sequences or reproduce time intervals often require experimental setups 
that allow recording response times with millisecond precision. Most setups
present auditory stimuli using either MIDI devices or specialized hardware such
as Arduino and are often expensive or require calibration and advanced
programming skills. Here, we present in detail an experimental setup that only
requires an external sound card and minor electronic skills, works on a
conventional PC, is cheaper than alternatives and requires almost no
programming skills. It is intended for presenting any auditory stimuli and
recording response times with within 2 milliseconds precision (up
to -2ms lag). This paper shows why desired
accuracy in recording response times against auditory stimuli is difficult to
achieve, presents an experimental setup to overcome this and explains in detail
how to set it up and use the provided code. Finally, code for analyzing input
recordings was evaluated, which shows that no spurious or missing events were
found in 94\% of the analyzed recordings.
}

\begin{document}
\setlength{\parindent}{0.5in}
\maketitle

Humans have a very distinct ability to synchronize motor movements to regular
sound patterns. We can finger-tap or sway along to a metronomic pulse and,
moreover, we are able to extract an underlying clock, the beat, from
non-isochronous rhythmic patterns \citep{repp2013sensorimotor}. Beat perception
is a fundamental component for experiencing music, ranked among life's greatest
pleasures \citep{dube2003content}. 

This ability to synchronize movement to an external stimuli \textemdash known
as sensorimotor synchronization or SMS \textemdash has been studied in detail.
Studies have revealed slowest and fastest tapping rate limits, what is the most
common spontaneous tapping rate and how it evolves from faster to slower with
age, that age allows us to synchronize to a wider rate range and that musical
training improves synchronization accuracy. Several models of how we
synchronize to rhythms and perform corrections in our tapping to compensate for
changes in the pacing signal have been introduced and tested experimentally.
When analyzing this phenomenon from the perspective of music, studies have
found that the rhythmic structure of the musical signal affects synchronization
precision. For a full review, please refer to \cite{repp2006musical} and
\cite{repp2013sensorimotor}.

\begin{figure}
\caption{\\ 
\textit{Schematic of Trial in a Sensorimotor Synchronization (SMS) Experiment.}
}\label{fig:trial}
\includegraphics[width=\textwidth]{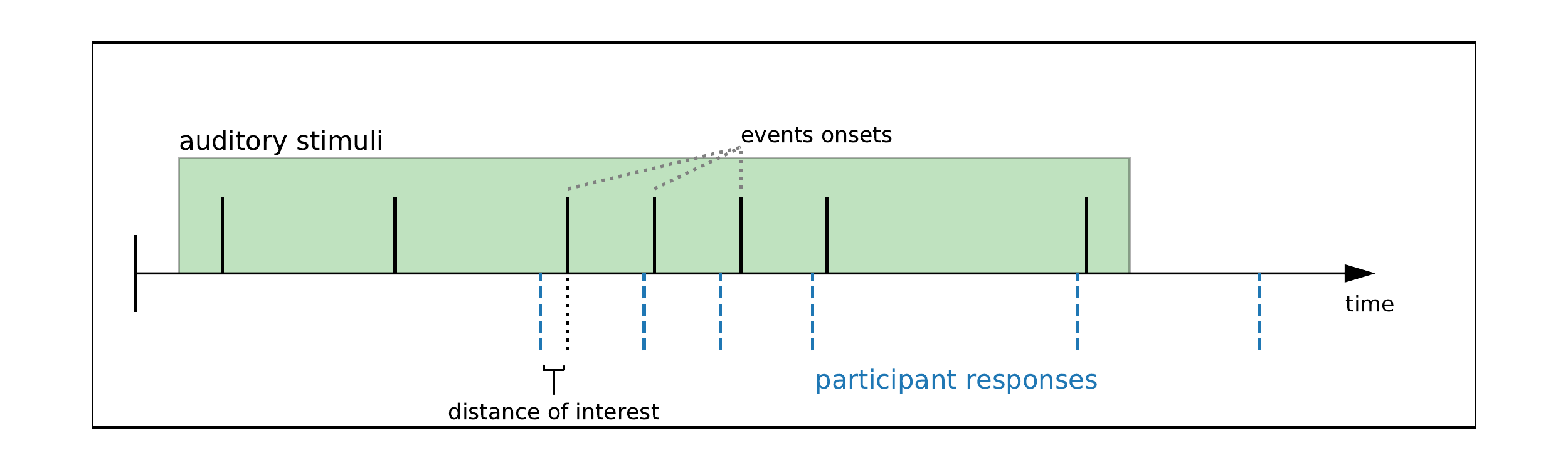}
\fnote{A trial consist of an auditory stimuli with identifiable
onsets developing in time. A participant has to listen to the stimuli and
produce responses. The measure of interest is the time interval between the
participant's response and the stimulus' onset time.}
\end{figure}

To understand these phenomena, behavioural studies require an experimental
setup that allows presenting an auditory stimuli and record participants' 
responses with great time fidelity. In several cases, it is also important to
capture the asynchrony between a participant response and the onset times
present in the stimulus. Figure \ref{fig:trial} presents the common scheme of a
trial in an SMS experiment.

\begin{figure}
\caption{\\
    \textit{Instantiations of SMS Trials in Various Experiments.}
}\label{fig:exp-examples}

\subfloat[Schematic of auditory trial in \cite{krause2010perception}. Participants 
tapped in synchrony to the metronome while it was being heard.
\label{fig:exp-examples:a}]{\includegraphics[width=0.5\textwidth]{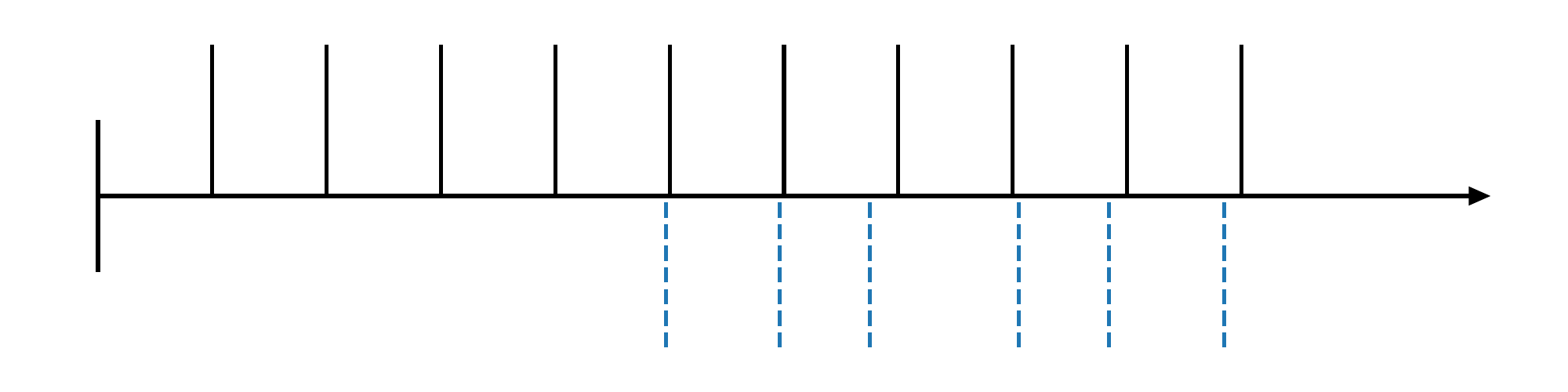}}
\subfloat[Schematic synchronization-continuation task in \cite{mcauley2006time}. Participants 
tapped in synchrony to the metronome while and after it sounded.
\label{fig:exp-examples:b}]{\includegraphics[width=0.5\textwidth]{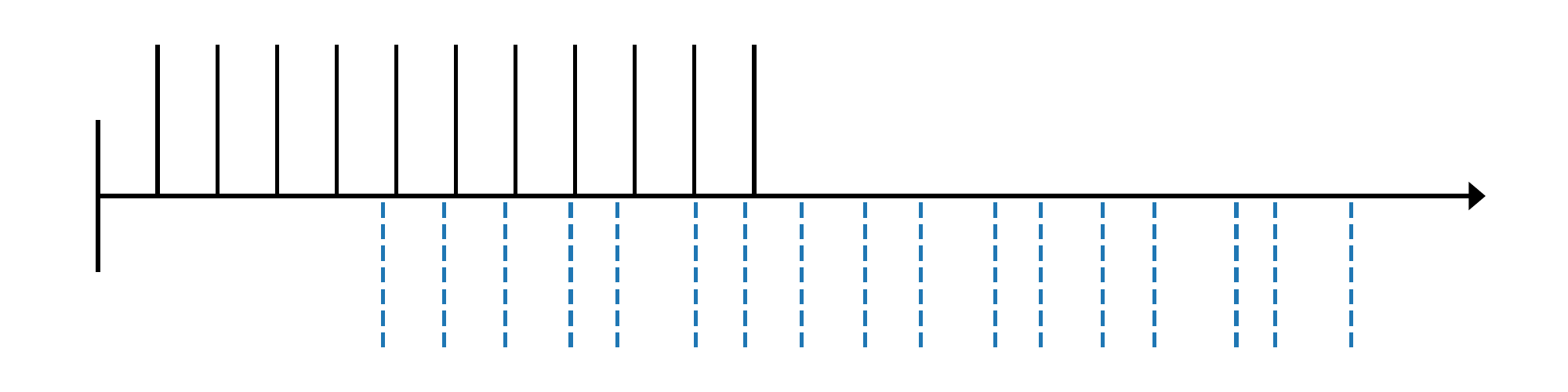}}

\hfill
\subfloat[Schematic synchronization-continuation task in \cite{repp2005production}. Participants 
tapped in synchrony to the rhythmic pattern and then reproduced it while a
metronome sounded. \label{fig:exp-examples:c}]{\includegraphics[width=0.5\textwidth]{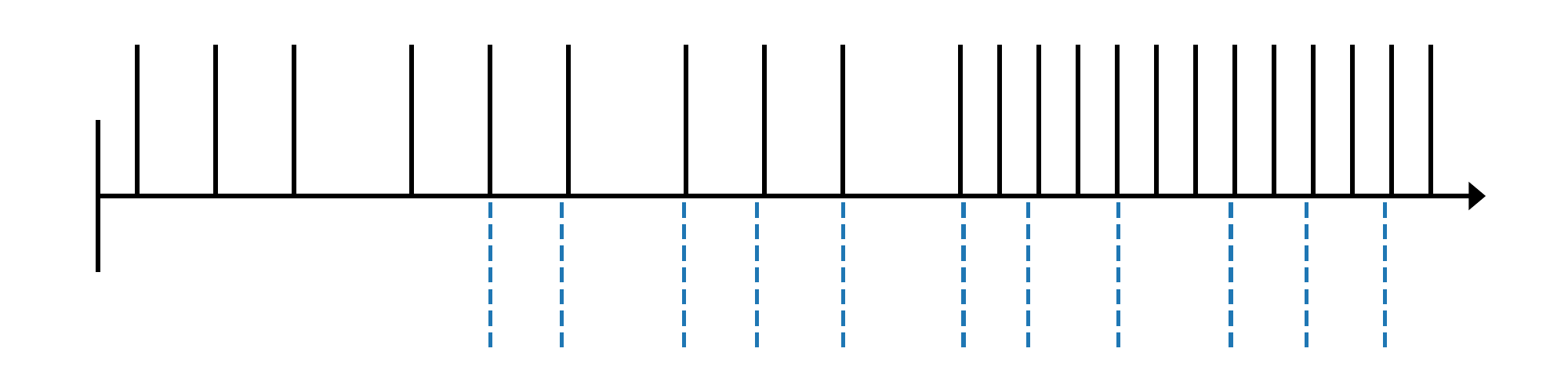}
}
\subfloat[Schematic Go/No-Go task in \cite{barry2014sequential}. Participants 
were asked to tap as soon as they heard the target stimulus or ignore it
otherwise. \label{fig:exp-examples:d}]{\includegraphics[width=0.5\textwidth]{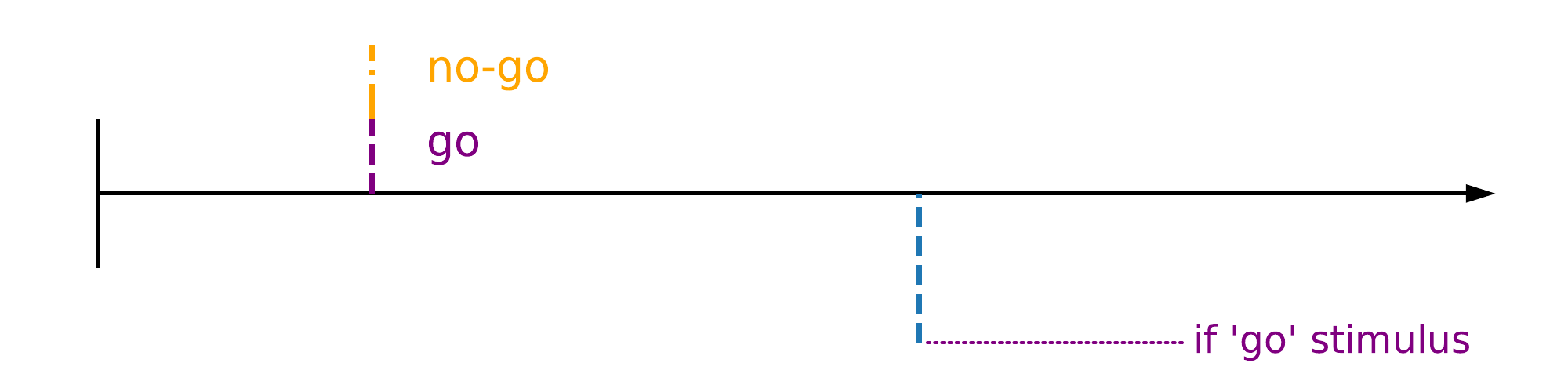}
}
\hfill
\subfloat[Schematic time interval reproduction task in \cite{daikoku2018motor}. Participants 
were asked to reproduce the heard time interval by performing to taps.
\label{fig:exp-examples:e}]{
    \includegraphics[width=0.5\textwidth]{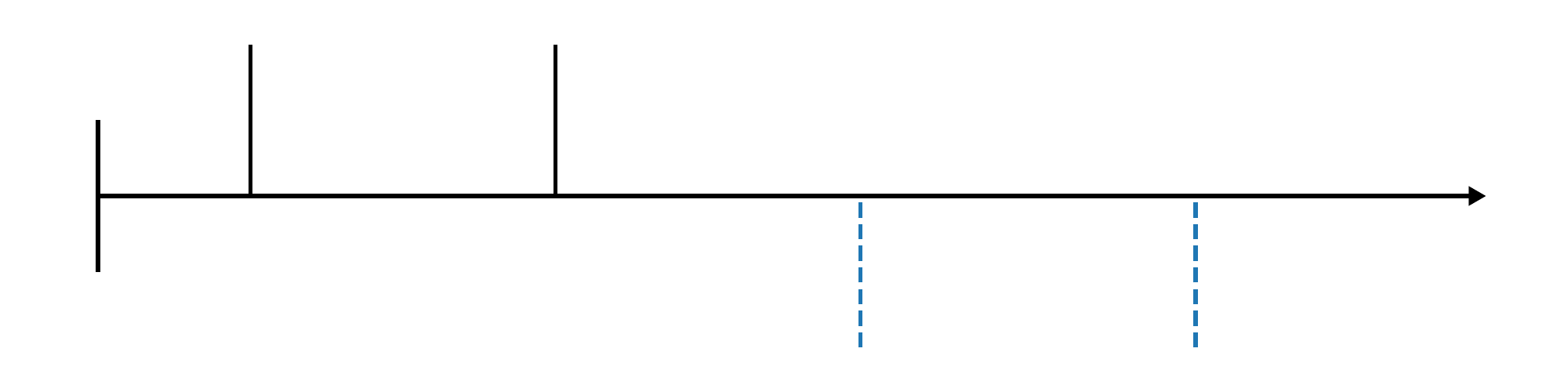}
}

\end{figure}

This general scheme can be instantiated in experiments performed in the 
literature, as presented in figure \ref{fig:exp-examples}. The study in
\cite{krause2010perception} explored the relationship between sensorimotor
synchronization and musical training. One of the tasks consisted of tapping in
synchrony to an isochronous stimuli in two modalities: visual and auditory. In
the auditory modality, the trial presented an isochronous tick to which the
participants had to synchronize until it stopped (see figure \ref{fig:exp-examples:a}).
In \cite{mcauley2006time}, participants of different ages were asked to tap in
synchrony to a metronome to study whether age changed the ability to synchronize
at different tapping rates. Trials started with an isochronous
tick to which the participant had to synchronize, but they were also asked to
continue tapping to the metronome's rate after it had stopped. This paradigm is
known as synchronization-continuation (see figure \ref{fig:exp-examples:b}).
The task presented in \cite{repp2005production} asked participants to
synchronize to non-isochronous stimulus by reproducing it. It also uses a
synchronization-continuation paradigm where the continuation phase may contain
a pacing signal instead of the original signal (see figure
\ref{fig:exp-examples:c}). 

Other paradigms that fit into the general experimental setup scheme proposed
are auditory Go/No-Go \citep{barry2014sequential} tasks and auditory time
interval reproduction \citep{daikoku2018motor}. The Go/No-Go task presents one
of two stimuli, with one designated as \emph{target}. During the
experiment, each trial consists of presenting one of the possible stimuli and
participants must respond only when the target stimuli is presented. In the
auditory mode, stimuli are sounds, and target stimulus may be distinguished,
for example, by pitch. In a time interval reproduction task, a time interval is
presented by two sounds separated in time. Afterwards, participants must try
to reproduce the interval as accurately as possible. Both task schemes are
presented in figures \ref{fig:exp-examples:d} and \ref{fig:exp-examples:e},
respectively.

Recording stimuli onset times and participants' response times
precisely cannot be directly achieved in an experimental setup using only a
computer and require specialized equipment or software. Setups presented in the
literature either use specific input devices (mostly MIDI instruments)
\citep{snyder2001tapping, fitch2007perception, repp2005production,
patel2005influence}, specialized data acquisition devices
\citep{elliott2014moving} or low-level programming of a microcontroller (e.g.:
Arduino) to work as an acquisition device \citep{schultz2019roles,
bavassi2017sensorimotor}.  Drawbacks of these setups come either from the cost
of the equipment or the technical skills required. MIDI input devices and data
acquisition devices (DAQs) generally used cost over 200 USD. A programmable
micro-controller is cheaper (about 30 USD) but requires low level programming
skills and does not include the input device.

In this paper we present an experimental setup that requires no programming
skills, is simple to assemble and costs under 60 USD (including the input
device). Beyond simplicity and
affordability, the setup proposed focuses on reliably capturing the time
interval between stimulus onset and the participant's response. To manage this
using MIDI devices, latency times for both input and
output devices must be verified \citep{finney2016defense}. On the other hand,
using a micro-controller can provide accurate stimulus timing but
cannot easily produce rich sounds \citep{schultz2016tap}. More recently,
some software setups allow presenting auditory onsets with precision and ease,
but still rely on expensive input equipment to gather timely responses
\cite{bridges2020timing}.

The next section ({\bf \nameref{sec:problem_description}}) describes in detail why
it is not straightforward to record stimuli to response time intervals using
only a computer's input and output hardware and more sophisticated 
solutions are required. This section also reviews previous solutions to the
problem. In section {\bf \nameref{sec:setup}} the setup presented here is described
in detail along with assembly instructions. The main features and limitations
of the setup are described in detail there. The section {\bf \nameref{sec:usage}}
presents the software tools provided to use the hardware setup and the
evaluation performed on the precision of how participants' responses are
collected.
 
\section{Problem Description}\label{sec:problem_description}
The common expected setup to run the described experiments in a personal
computer would be to have the computer present the stimuli (in this case,
audio) and simultaneously record participants' responses via standard input
devices such as a keyboard or a mouse. In more detail, this situation implies
the following steps in time:

\begin{enumerate}
\item Computer program produces auditory stimuli and records stimuli onset time
($o^c$).
\item Sound is produced on the speakers or headphones and heard by the
participant ($o^p$).
\item Participant produces a response by operating the input device (keyboard,
mouse, etc.) ($r^p$).
\item The response is captured by the computer and response time is recorded
($r^c$).
\item Response time is calculated from the times recorded by the computer
($rt^c = r^c - o^c$).
\end{enumerate}

\begin{figure}
\caption{\\ \textit{Depiction of the Delays Between Computer and Participant's Stimuli
Onset and Response Times.} }\label{fig:delays}
\includegraphics[width=0.9\textwidth]{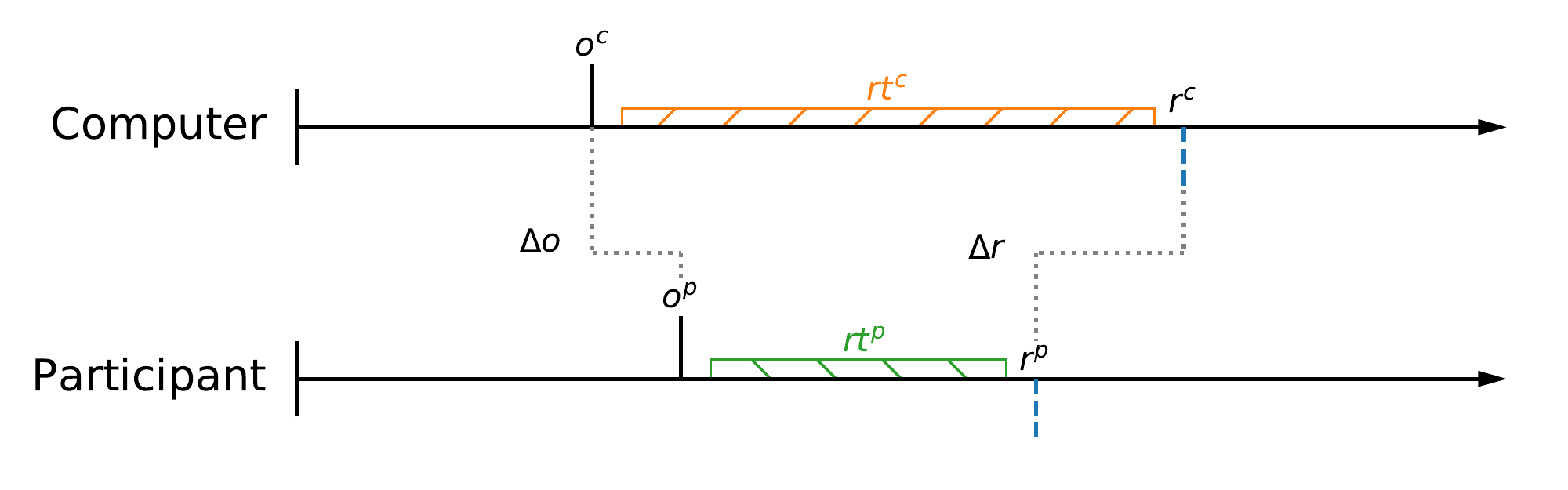}
\fnote{In a common computer setup, response times are
calculated from the onset and response times known to the computer ($o^c$ and
$r^c$, respectively). These times may differ from the actual onset and response
times perceived and produced by the participant ($o^p$ and $r^p$). As a result,
the obtained response time ($rt^c$) is different from the one that
is of interest ($rt^p$).}
\end{figure}

The situation described above is depicted in Figure \ref{fig:delays}. The issue
with this simple conception of the experimental setup is that the 
moment where the auditory stimuli is actually produced may be significantly
delayed from the moment the computer program decided to produce the sound
($\Delta o = o^p - o^c$).
Additionally, the time the computer learns a keyboard key is pressed can be
much later than the time the key was effectively pressed ($\Delta r = r^p -
r^o$). As a consequence, the response time captured ($rt^c$) may be different
than the real response time ($rt^p$).

The delays in stimuli production ($\Delta o$) and response capture ($\Delta r$)
are analyzed in two magnitudes: lag (or accuracy) and jitter (or precision).
Lag refers to a constant delay between the two onset or response times. Jitter
refers to the unknown variability in the delay. If the delay is thought of as a
random variable, the lag or accuracy would be represented by the expected delay
and the jitter or precision with the standard deviation. Depending on the
setup and experimental question at task, lag can be cancelled out. In some
setups it may be possible to measure and subtract or, if the question is a
comparison between groups, the comparison of response times will inherently
ignore such delay. On the contrary, jitter is unknown and my affect differently
each trial, being therefor more difficult to disregard.

In common computers, onset delay ($\Delta o$) may be caused by several reasons.
In a standard installation with an operating system, several layers of software
drivers exist between the experiment's program and the sound card that
translates the digital encoding of sound into electric pulses for the speakers.
These layers may cause the message to be delayed between the user's software
and the hardware. The conversion between digital and analogue representations
of sound, made by the digital-to-analogue converter (DAC), also requires time.
Finally, some sound producing devices, such as MIDI instruments, may also
require time to process the onset digital signal to effectively produce sound.
Regarding responses, capture delays ($\Delta r$) can also be product of
the transit from the driver receiving information from hardware devices and
the experiment's software. Some devices may also introduce delays between
receiving the participant's pressure action and producing a signal. For
example, standard keyboards are known to have a lag larger than 10 ms, with
varying jitter of about 5 ms \citep{segalowitz1990suitability,
shimizu2002measuring, bridges2020timing}.

\subsection{Proposed approaches}

There are two main approaches to overcome the latency problems described:
either reduce the delays ($\Delta o$ and $\Delta r$) below a required value or
record the actual onset and response times perceived and provided by the
participant in a way that is independent of the computer and does not
introduce relevant latencies.

\cite{finney2001ftap} takes on the first approach and uses MIDI devices for
both input and output. The Musical Instrument Digital Interface (MIDI) is a
communication protocol for sending and receiving information on how and when to
play musical notes through MIDI devices. Examples are physical instruments such
as keyboards or drum pads that can produce messages, synthesizers that receive
MIDI messages and produce sounds or computers with MIDI ports that may do both. In
his work, Finney presents FTAP, a software tool for running experiments
involving auditory stimuli and response time collection. For timing precision,
FTAP takes advantage of the MIDI protocol's capacity to exchange messages at
approximately one message per millisecond. The software package includes an
utility to test whether such exchange rate is achieved in a specific computer
setup (computer, operating system, MIDI drivers and MIDI card) as not every
configuration allows such optimal rate. Moreover, testing of the input and
output hardware used is recommended, as it has been seen that some MIDI input
devices can introduce delays of several milliseconds from the moment it is
actuated until the MIDI message is sent \citep{schultz2016tap,
finney2016defense}. End-to-end testing of an experimental setup implies
measuring the complete time from the participant's input (pressing on the
device) to the time the computer captures the message or auditory feedback is
produced, depending on the requirements of the experiment. Commercial equipment
for this purpose has been presented in \cite{plant2004self} and an alternative
using an Arduino controller is described in \cite{schultz2019schultz}.

\cite{schultz2016tap} focuses on presenting a setup to provide timely auditory
feedback to a participant's response. To do so, they capture the response and
provide feedback using a programmable micro-controller (namely Arduino).
Micro-controllers often provide input and output pins that allow interfacing
with external hardware by measuring and producing changes in the voltage of the
electrical current that runs through the pins. The importance of using a
microcontroller lies in that it provides the programmer with direct access to
the processor and the input and output pins. This allows more precise control
over the delays of processing the input and producing the output in comparison
with using a computer with an operating system where the delays of the hardware
drivers and those introduced by the multitasking capabilities are harder to
manage or know.
In their proposed setup, they
capture the participant's tap using a Force Sensitive Resistor (FSR)
(depicted in figure \ref{fig:fsr_setup}). An FSR is a device shaped as a flat
surface that varies the voltage of an electrical current according to
the pressure it receives.  Such changes in voltage can be measured in an input
pin in the micro-controller to recognize when the device is being pressed.
Finally, they test two methods to provide feedback from the Arduino. One
method is connecting a headphone directly to an output pin of the controller.
This allows the program to produce feedback very quickly (delay of 0.6 ms, sd
of 0.3 ms) with the caveat of it being a simple sound. Another method tested is
the Wave Shield for Arduino, an extension hardware that allows reproducing any
sound file. The Wave Shield feedback requires more time to emit a sound (2.6 ms
in average) but still has low jitter (0.3 ms).

The timing mega-study in \cite{bridges2020timing} analyzes onset lag and jitter
for auditory and visual stimulus on a variety of existing software packages for
designing and running behavioral experiments on PC. These packages provide
utilities to generate programs that run experiments, present stimuli,
collect answers, randomize trial conditions, among other
features commonly used. Moreover, these software packages manage drivers and
settings in order to produce onsets, both auditory and visually, with less than
a millisecond unknown delay. The results of this work shows that a common
computer hardware can be used to produce auditory stimuli quickly in spite of
the stack of drivers and multitasking mentioned. To do so, the right software
configuration is required. For example, PsychoPy must be updated to version
3.2+ which recently included the correct software to achieve millisecond
auditory onset presentation. The issue still remains on the precision of the
input capture, which in \cite{bridges2020timing} is solved by using a
specialized response button device.

The MatTAP tool suit, presented in \cite{elliott2009mattap}, uses the second
approach to work around the delays and achieve precise recording of stimuli and
response times. This approach is based on using a recording device independent
of the computer running the experimental procedure in such a way that producing
the stimulus and recording responses is not affected by the software stack
of the operating system. More specifically, they use a Data Acquisition device
(DAQ). DAQs are devices that can produce and record multiple analogue and
digital signals simultaneously with a sampling rate of hundreds of kilo-samples
per second, providing sub-millisecond precision. The MatTAP tool suit is a
MATLAB tool box that communicates with the DAQ in order provide the stimuli
onset times and sounds. The DAQ then produces the sounds and simultaneously
records the input from the input device. Each auditory onset is accompanied by
a digital onset on a separate channel that is required to be looped back into a
recording channel of the DAQ. Although it produces the stimulus without
lag relative to the stimuli sequence, there may be a delay introduced by the
initial communication between the computer and the DAQ. As a consequence, the
loop back is required to be able to synchronize the stimuli onset times with
the response times (see the next section and fig \ref{fig:rec_approach} for a
more detailed explanation). Finally, if the DAQ has more than 2 input channels,
MatTAP is capable of recording two input devices. Also, the digital output
signal produced with each stimulus onset may be used to drive another output
device. The MatTAP tool suit provides software utilities for creating up to
two metronomes for synchronization experiments, allows managing settings for
multiple trials and also collects the experiment response data for each trial.
The tool suit also provides a customizable utility to analyze the input
signals, extract responses and calculate stimulus to response asynchrony times.

The setup proposed here also follows the second approach. In comparison with
MatTAP \citep{elliott2009mattap}, we use an external sound card as an
acquisition device, which works as a less expensive replacement. Moreover, the
software tool set provided here does not require proprietary software such as
MATLAB. Finally, we present instructions to assemble an input device that
can be connected to the sound card and provide accurate response time
recording.
 
\section{Setup Description and Installation}\label{sec:setup}
In {\bf \nameref{sec:problem_description}} we established two main approaches
to solve the issues that arise when recording response times to auditory
stimuli due to delays in both onset presentation and response time collection.
One approach is to reduce such delays below an accepted value. Another approach
is to record the onset times actually perceived and produced by the participant
with an independent recording device. Our setup takes the second approach and
proposes to do so using either the sound card already present in common desktop
computers or an inexpensive external sound card. In this section we present why
our setup addresses the problem, how our setup is assembled and what is the
expected workflow for running experiments with it. To complete the setup, this
work presents instructions to assemble a pressable input device using a force
sensitive resistor (FSR) and an open-source software tool suit to produce the
stimuli, record the responses and obtain response times from our proposed input
device.  The instructions to assemble the input device are introduced in the
next subsection. Then we present instructions to connect the input device, the
recording device and test the setup connections. The tool suit is introduced in
the next section. 

\begin{figure}
\caption{\\ \textit{Representation of Stimuli and Recording Times when Using an
Independent Recording Device.}}\label{fig:rec_approach} 
\includegraphics[width=\textwidth]{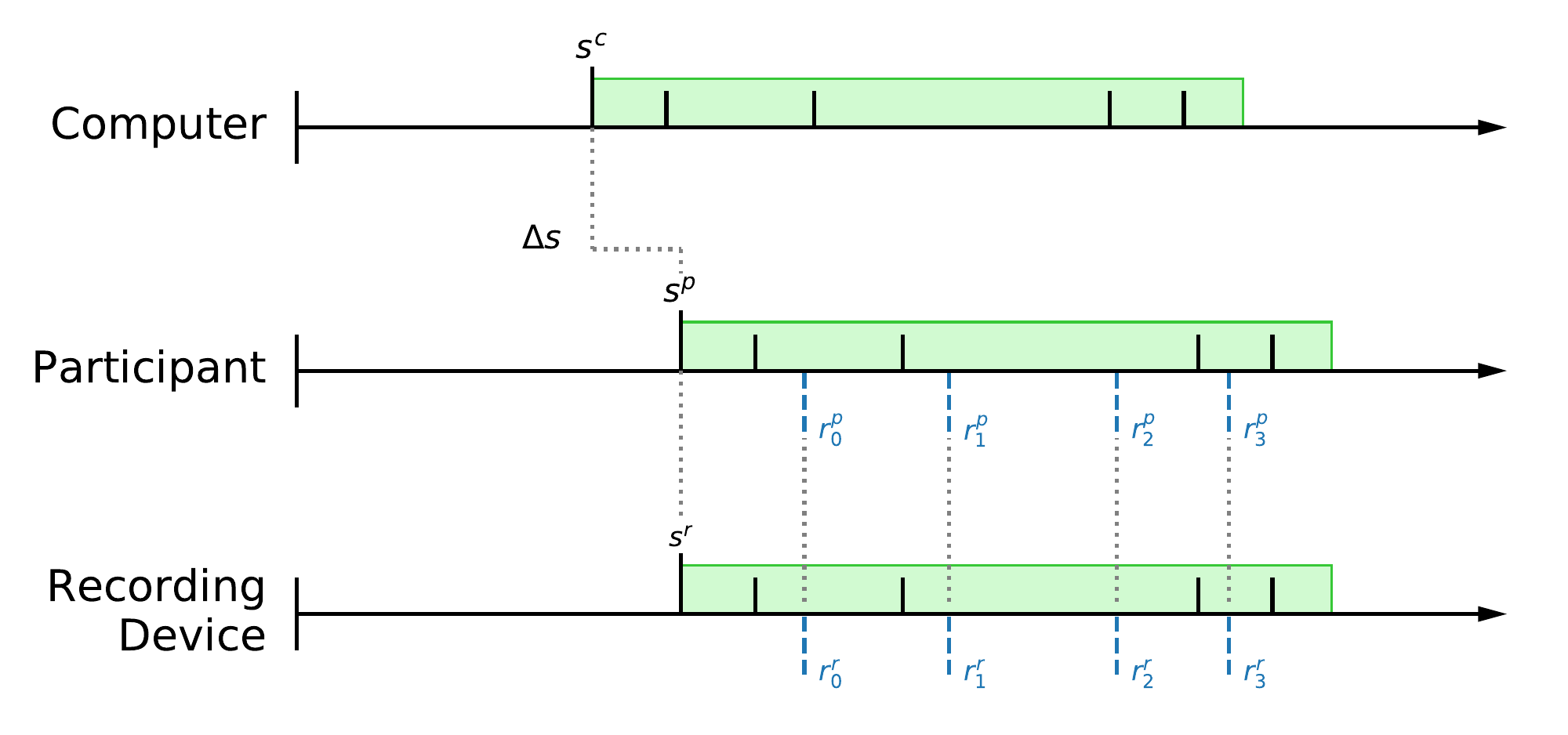}
\fnote{This approach adds a new device that
simultaneously records the stimuli and the responses with high accuracy.
Stimuli are presented as one audio with multiple onsets (box containing
lines). The stimulus presentation time may lag from the computer command but is
recorded at the same time it is heard by the participant ($s^c \ge s^p$ and
$s^p = s^r$). Response times ($r_i$) are captured in synchrony with stimulus
presentation. }
\end{figure}

The key component of the approach used in this setup is to be
able to record both the participant's responses and the stimuli simultaneously. 
This allows having the stimuli synchronized with the responses on one device's
timeline. Because one of the delays is introduced between the moment the
experiment code produces a stimulus and when it is played on the output device,
producing each stimulus separately would add variablity to the inter-stimuli
interval. Such behaviour can render certain experimental conditions unusable.
Our proposal is to package all stimuli onsets into one audio file. This
introduces only one delay on when the whole stimuli set is reproduced ($\Delta
s = s^c - s^p$) but no variablity in inter-stimuli intervals. Finally,
stimuli and participant's responses are recorded by the device. Stimuli onset
times can be retreived by synchronizing the output audio file with the
recording. Response times can be obtained from the recorded signal relative to
the beginning of the stimulus.  How this procedure is performed is explained in
section {\bf \nameref{sec:usage}}. The setup and new definition of delays is
depicted in figure \ref{fig:rec_approach}.

\begin{figure}
\centering
\caption{\\ \textit{Schematic of the Connections of the Setup. }}
\label{fig:connections}
\includegraphics[width=0.8\textwidth]{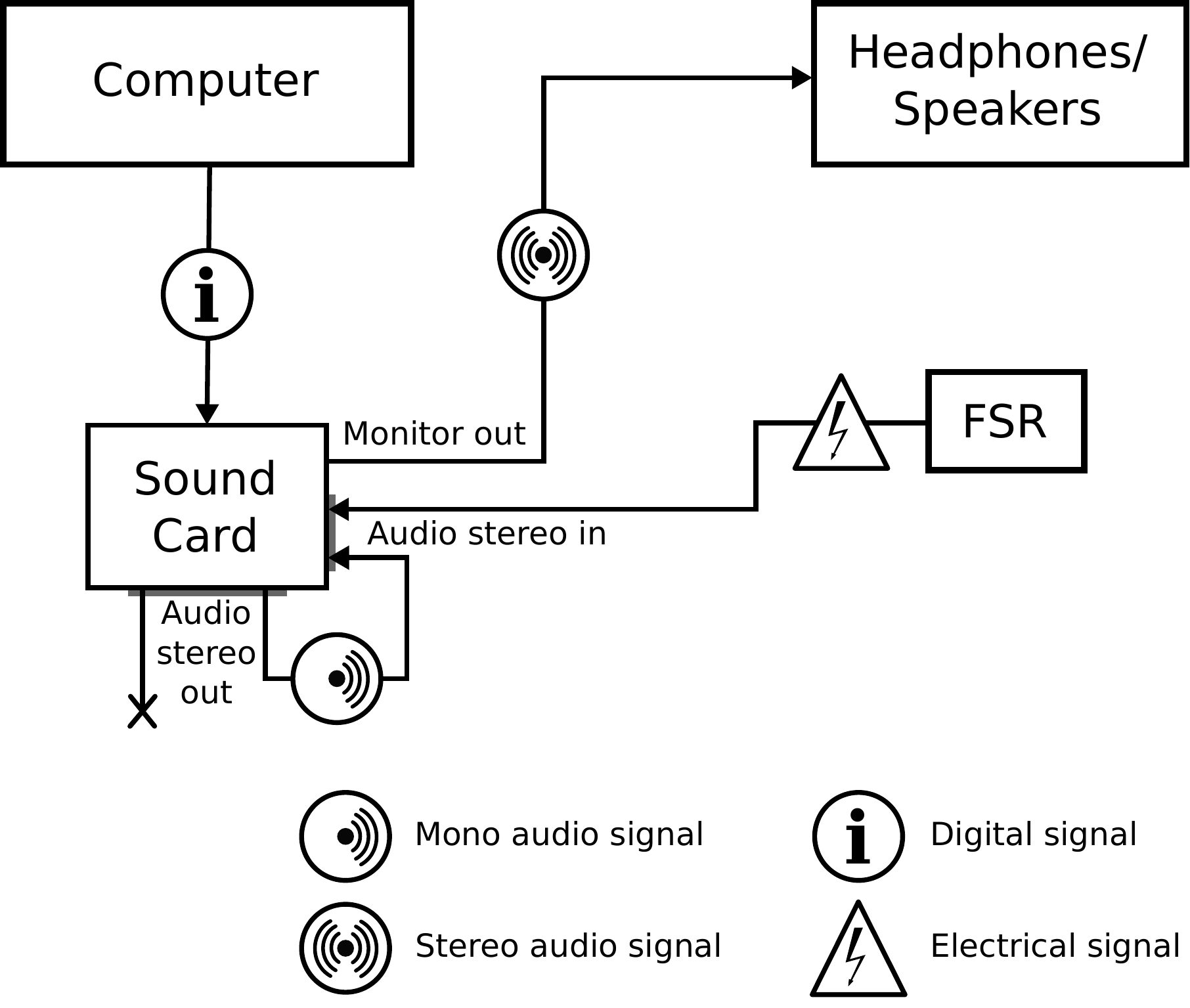}
    \fnote{Computer
exchanges information with the recording device (sound card). One of the
device's sound output is sent to the participant, the other has one channel
looped-back to one of the recording device's input channels. The response 
device is connected to the other input channel.}
\end{figure}

\begin{figure}
\caption{\\ \textit{Example of Stimulus Audio Signal (middle) and an Input Recording
(bottom).}}\label{fig:recording_input}

\subfloat[Stimulus onset times \label{fig:recording_signal}]{\includegraphics[width=\textwidth]{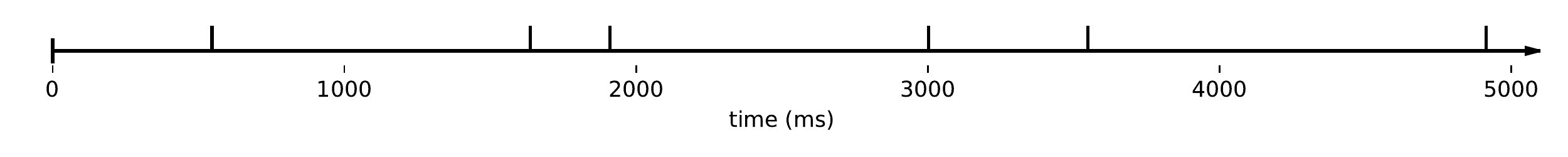}}
\hfill
\subfloat[Stimulus audio \label{fig:recording_output}]{\includegraphics[width=\textwidth]{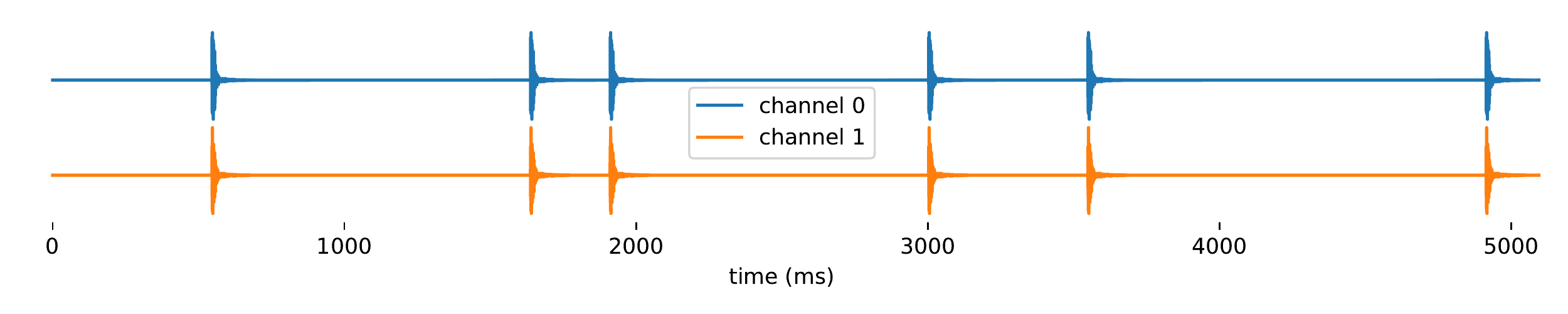}}
\hfill
\subfloat[Mixed loopback and input device recording \label{fig:recording_mix}]{\includegraphics[width=\textwidth]{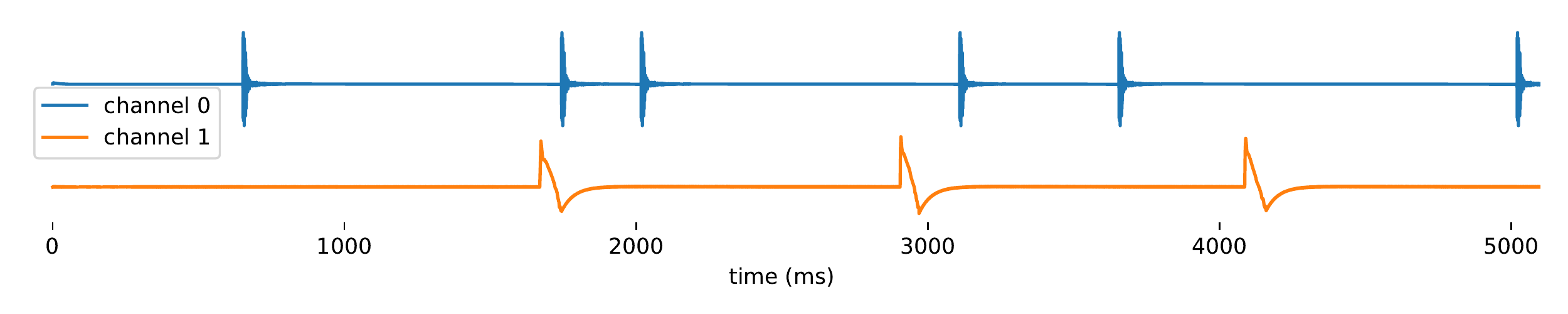}}
\hfill
    \fnote{The stimulus signal is an arbitrary stereo audio file
    \subref{fig:recording_output}. In this case, it is an audio signal
    constructed from designated onset times \subref{fig:recording_signal}.
    The input recording \subref{fig:recording_mix} contains on one channel
    the looped-back signal from the stimulus (0) and on the other the
electrical signal from the input device (1).} \label{fig:audios} 
\end{figure}

To achieve recording the stimuli and responses simultaneously, the recording
device used must have an stereo output and at least two input channels (or one
stereo input). With this, one channel of the stereo output can be looped back
into one of the input channels. Moreover, to keep the setup simple, our
proposed setup requires the recording device to have a secondary output that
mirrors the primary. While the primary output signal is looped-back into one
input channel, the secondary output is connected to the output device (speakers
or headphones). Finally, the signal of the response device is connected to the
second input channel of the recording device. With this connection setup, 
the audio input of the recording device can be collected into a stereo file
containing the stimulus in one channel and the response signal on the other.
The connections mentioned are depicted in figure \ref{fig:connections}. The
audio output and input signals are depicted in figure \ref{fig:audios}.

In the next subsection we present how to assemble the input device used
in the complete version of the setup. Then we show how to connect the input
device, the recording device and the computer, and then test that the
connections work correctly.

\subsection{Input device (FSR)}

\begin{figure}
\centering
\caption{\\ \textit{Setup Diagram for a Tapping Input Device Using a Force Sensitive
Resistor (FSR)}}\label{fig:fsr_setup} 
\subfloat[Schematic of the component connections]{
\includegraphics[width=0.9\textwidth]{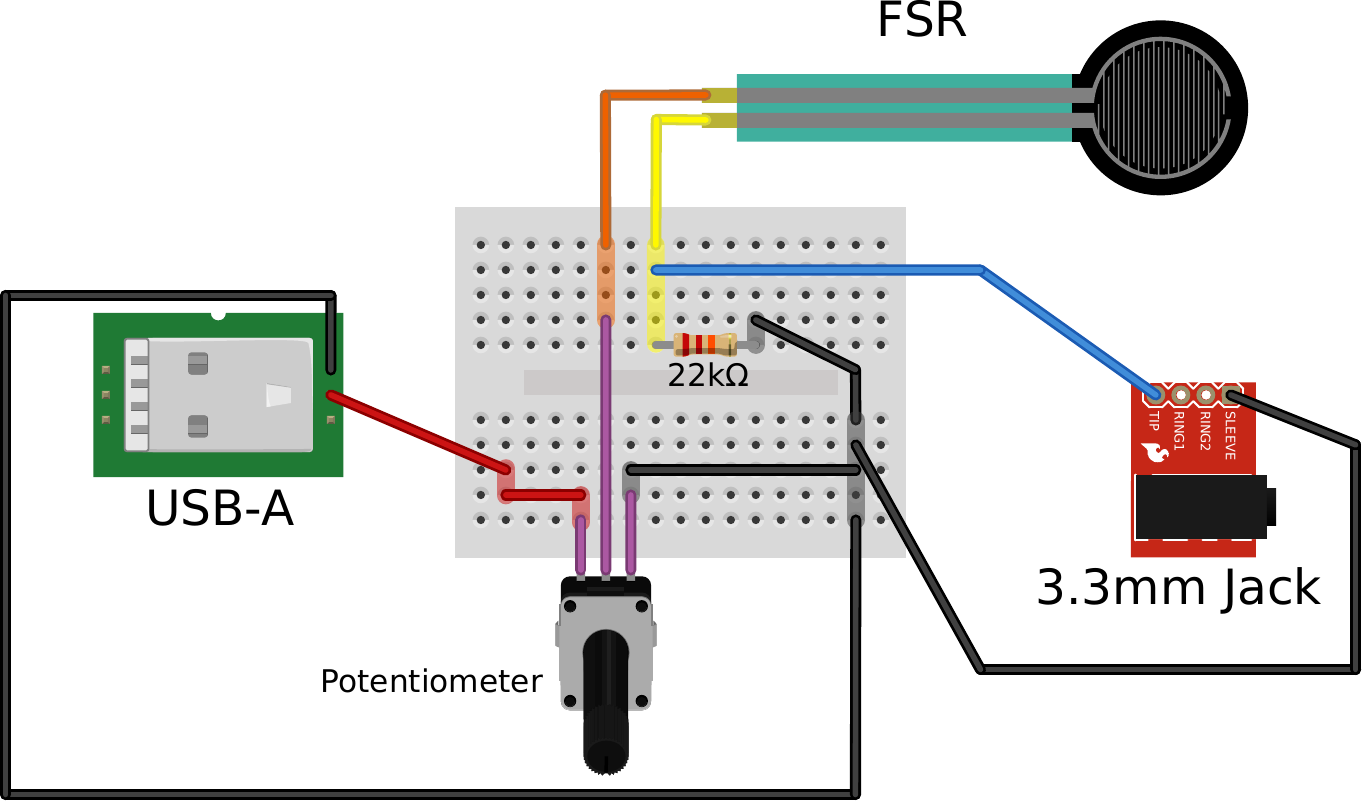}}
\hfill
\subfloat[Schematic of the electronic connections]{
\includegraphics[width=0.9\textwidth]{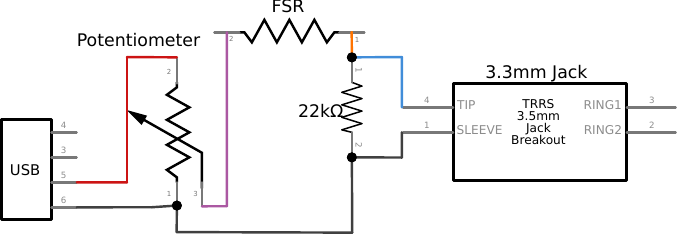}}
    \fnote{ 
    The setup drives current with varying voltage from a VBUS pin of a USB
    (type-a) connector to the pin of a 3.5mm audio jack. Voltage is limited
    using a voltage divisor circuit (orange and purple). Final output voltage
    into the audio jack depends on the resistance provided by the FSR, that
    drops with pressure. The higher the pressure, the higher the voltage
    provided in the audio jack and recorded by the sound card.
    Schematic was drawn using Fritzing \citep{knorig2009fritzing}.  }
\end{figure}

Our setup uses a pressable input device. The main component of the device is a
Force Sensitive Resistor (FSR), a flat sensor whose electrice resistance is
reduced when pressed (fig \ref{fig:fsr_setup}). The variance in
resistance can be used to create a variation in voltage that can be recorded by
the audio input of a sound card. The proposed device uses a 3.5mm female audio
jack that allows connecting the input device with the sound card using standard
audio cables. 

The circuit allowing this variation requires a
voltage source. We propose using a standard USB (type-a) cable connected to a
computer for this purpose. The voltage variation provided to the sound card
can saturate the recording, depending on the device's sensitivity. This
saturation can modify the activation profile captured from the FSR. The
proposed circuit adds a voltage divisor that allows limiting the maximum
voltage received by the sound card. In our proposal, we use a 10k$\Omega$
potentiometer, another adjustable resistor, that can be set by trial and error
to prevent the signal from saturating the recording.

The proposed circuit is presented in figure \ref{fig:fsr_setup}. 
For the voltage source we
use a standard USB (type-a) cable connected to the computer. USB
cables have 4 pins, two for data, one that drives a 5V signal (VBUS or VCC) and
a ground connection (GND). The VBUS pin (fig \ref{fig:fsr_setup}, red cable) is 
connected to the divisor circuit which centerpiece is the potentiometer
(fig \ref{fig:fsr_setup}, purple cable). The division goes to the FSR
(orange cable) and back to the ground (black cable). The other end of the FSR
(yellow cable) is connected to the 3.5 jack (blue) and simultaneously grounded
through a 22k$\Omega$ resistor (black). More detailed instructions for assembly
are presented as a video tutorial, linked in the 
\emph{\nameref{sec:open_practices}} section.

\begin{figure}[h]
\centering
\caption{\\ \textit{Setup Diagram for a Tapping Input Device
Used}}\label{fig:fsr_setup_used} 
\subfloat[Schematic of the component connections]{
    \includegraphics[width=0.9\textwidth]{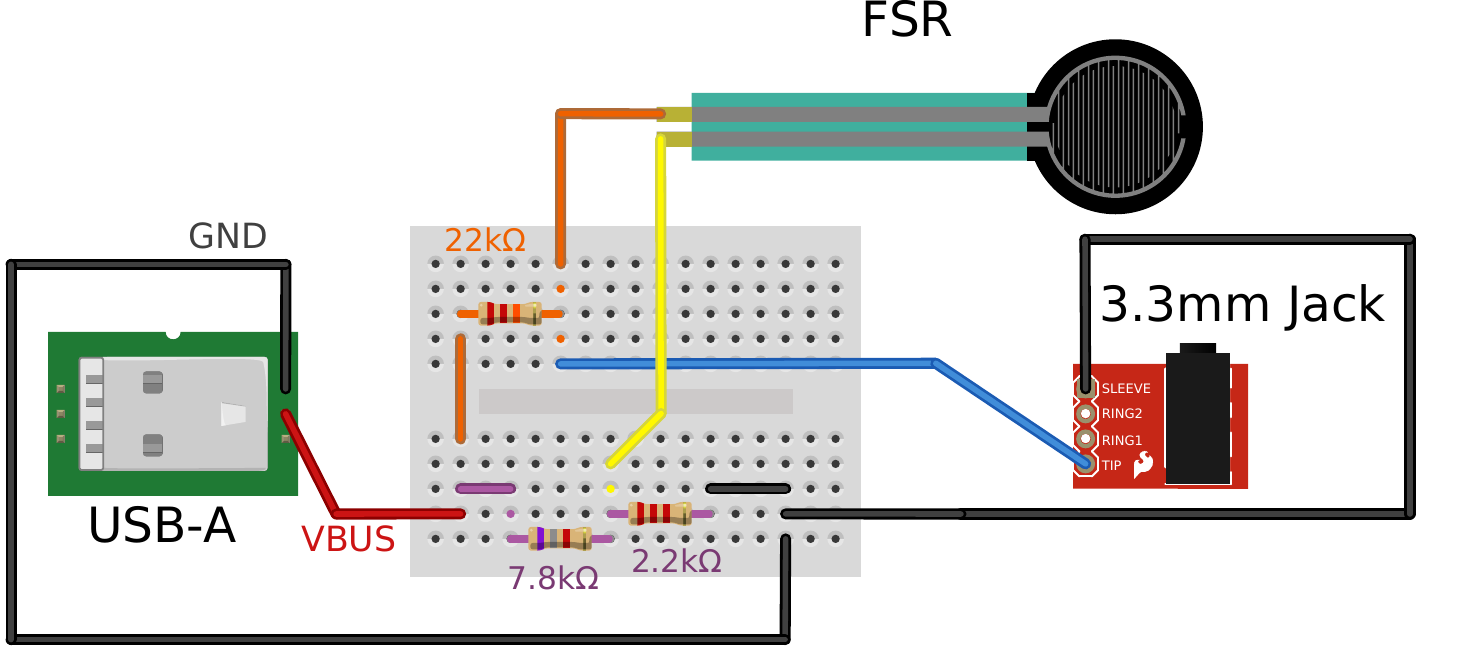}}
\hfill
\subfloat[Schematic of the electronic connections]{
    \includegraphics[width=0.9\textwidth]{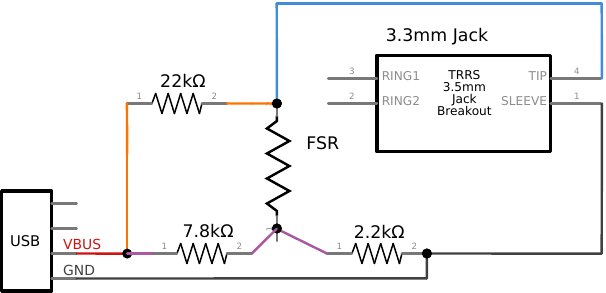}}
\fnote{Schematic was drawn using Fritzing \citep{knorig2009fritzing}.  }
\end{figure}

The next subsection explains how the FSR input device is connected with the
rest of the setup and how to test the connections. It also explains how to use
the potentiometer to adjust the signal amplitude to avoid saturation. 
\todo{Nombrar video?}
The exact circuit used in this work is presented in figure
\ref{fig:fsr_setup_used} and replaces the potentiometer with two resistors
selected for the sound card used (Behringer UCA-202). It has a further
adjustment to deliver a descending (instead of ascending) voltage change when
the FSR is actuated given that this sound card inverts the signal when
recording. The tool suit provided here has a setting that allows managing this
situation.

\subsection{Setup assembly}

\begin{figure}
\centering
\caption[position=top]{\\ \textit{Reference Pictures of the Elements and
Connections of the setup.}}
\label{fig:connections_fig}
\subfloat[Sound card with stereo input and output and monitor output \label{fig:sound_card}]{
    \includegraphics[width=0.3\textwidth]{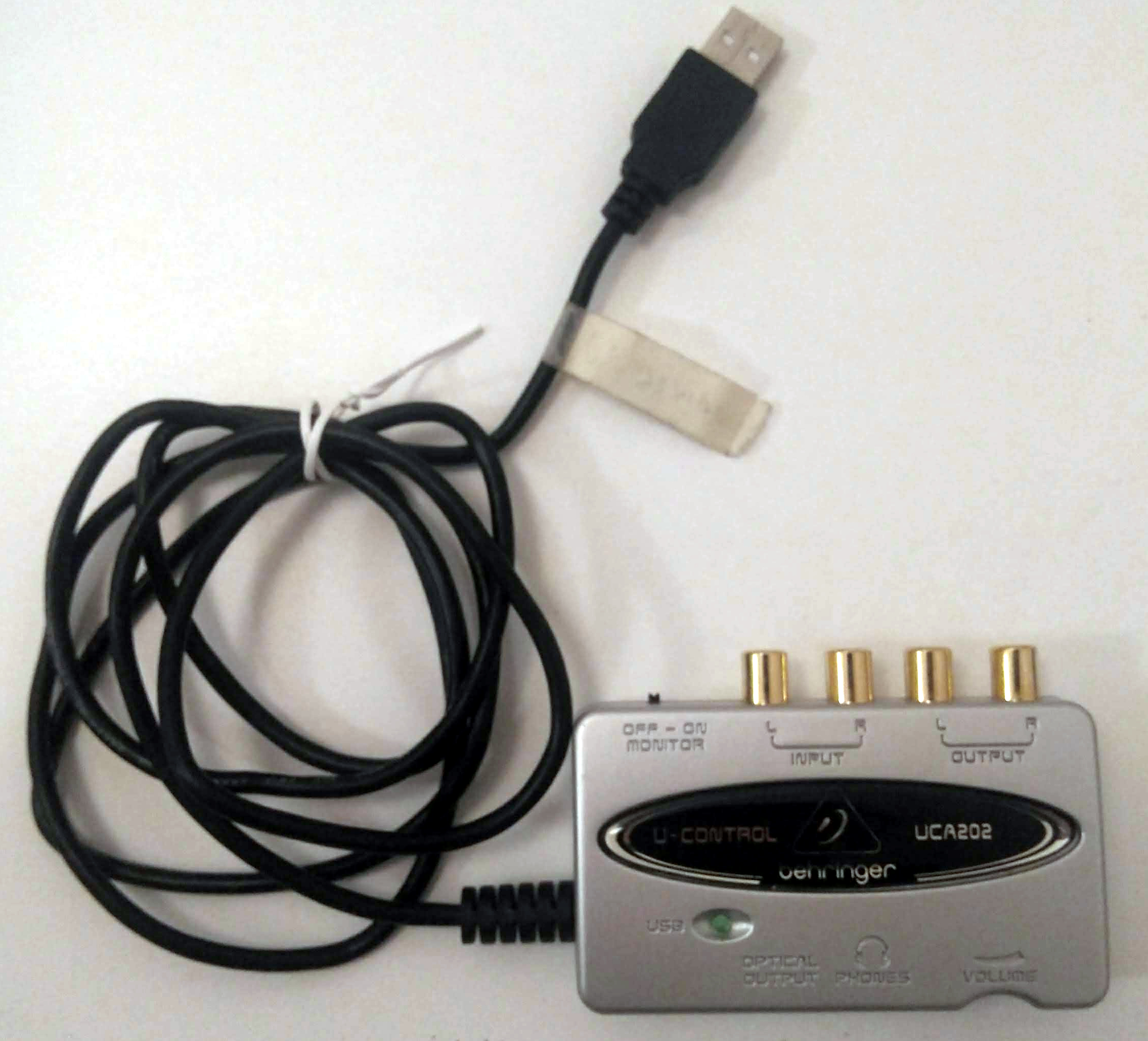}}
\begin{minipage}{0.04\textwidth}\end{minipage}
\subfloat[Stereo male-male rca cable. \label{fig:rca_rca}]{
    \includegraphics[angle=90,width=0.3\textwidth]{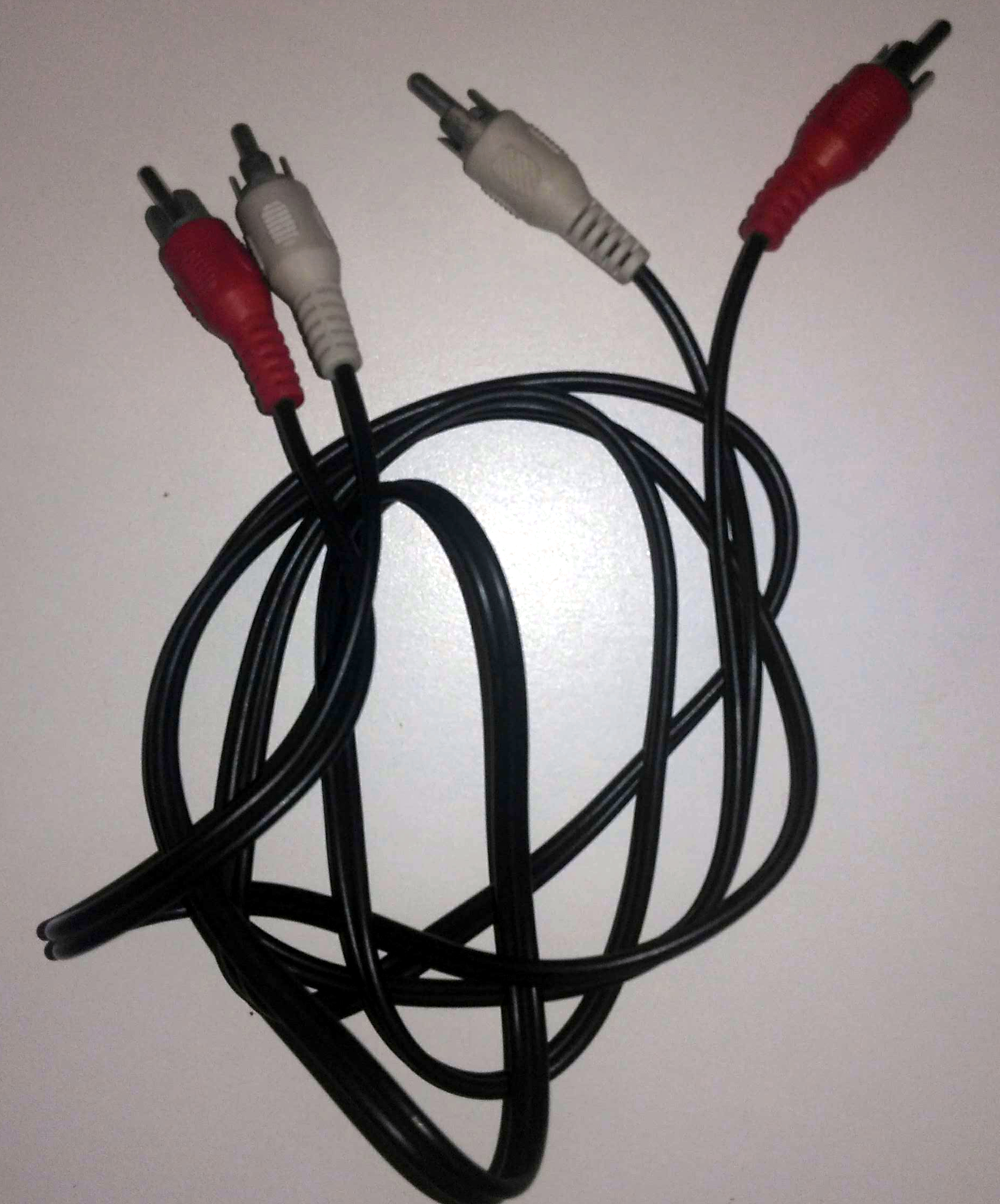}}
\begin{minipage}{0.04\textwidth}\end{minipage}
\subfloat[Stereo male-male rca-3.5mm jack cable. \label{fig:rca_plug}]{
    \includegraphics[width=0.3\textwidth]{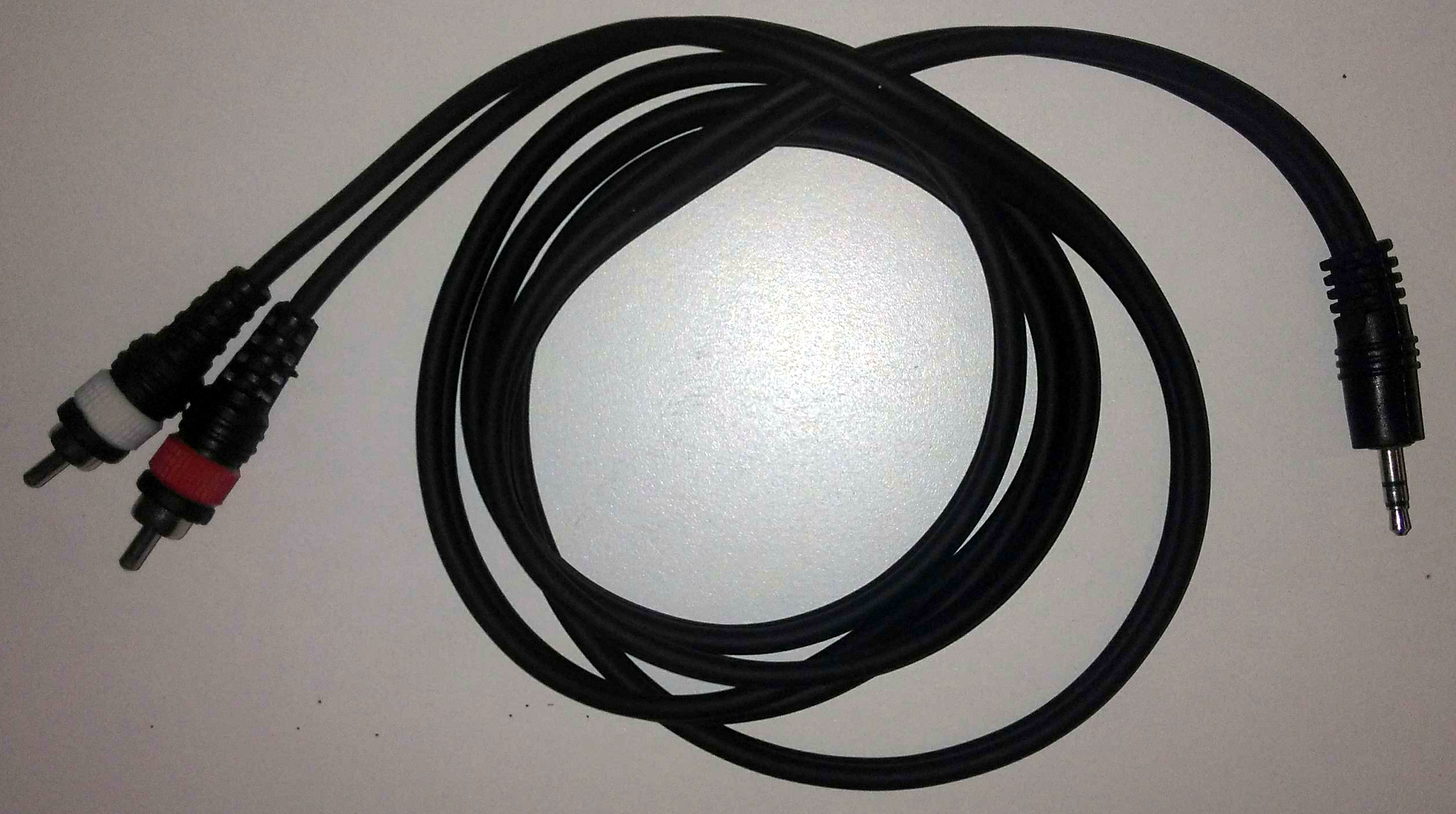}}
\hfill
\subfloat[Picture of connections in the setup. {\small It replicates the
diagram in figure \ref{fig:connections}.} \label{fig:connections_pic}]{
    \includegraphics[width=0.8\textwidth]{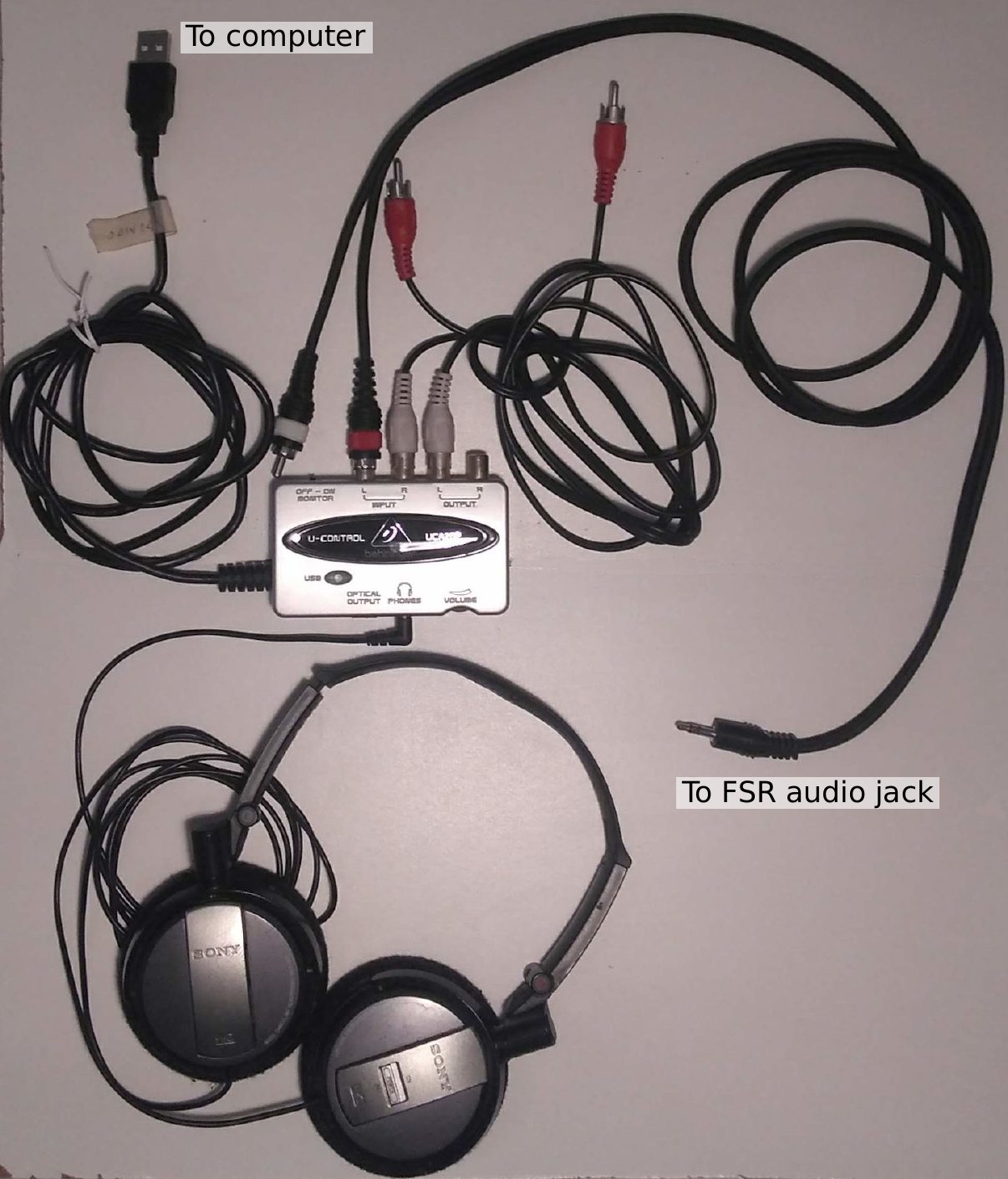}} 
\end{figure}

The three key components of the setup are the recording device with two input
channels and output channels, the loopback between output and input and the
mixing of the input device with the loopback in the stereo input. An schematic
of these connections is presented in figure \ref{fig:connections}. In figure
\ref{fig:connections_fig} we present the same connection scheme with the
picture of the devices connected. Given that our external sound card (fig
\ref{fig:sound_card}) uses RCA plugs, we use a male-male RCA-RCA cable to
perform the loopback (fig \ref{fig:rca_rca}). Although we used an stereo cable,
a mono cable is sufficient. Finally, we connect the FSR setup with the
recording device using a 3.5mm plug to RCA cable (fig \ref{fig:rca_plug}).
Again, we used an stereo cable, but a mono cable is sufficient.

\begin{figure}
    \centering
    \caption{\\ \textit{Interface of Audacity \citep{website:audacity} used to test the
    recording}}
    \subfloat[Relevant buttons and configuration]{
        \includegraphics[width=0.6\textwidth]{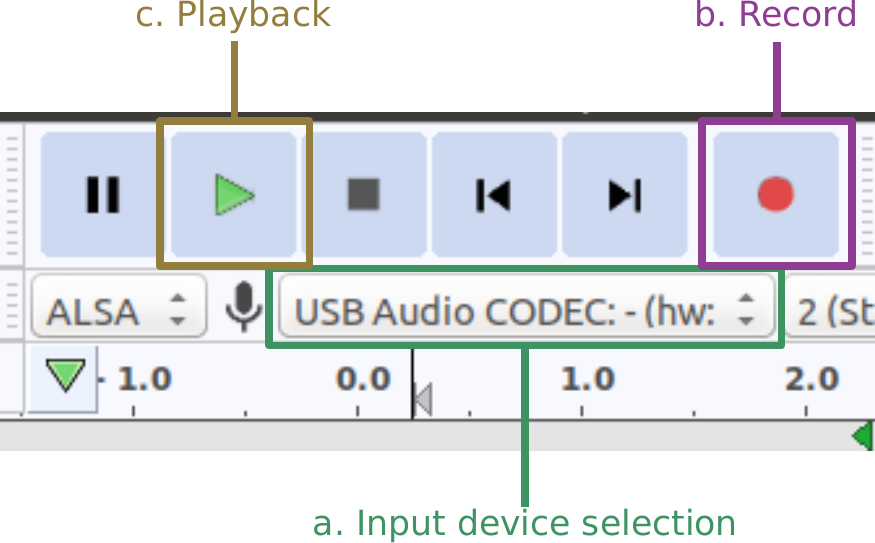}}
    \hfill
    \subfloat[Depiction of a recorded fsr peak \label{fig:recorded-peak}]{
        \includegraphics[width=0.5\textwidth]{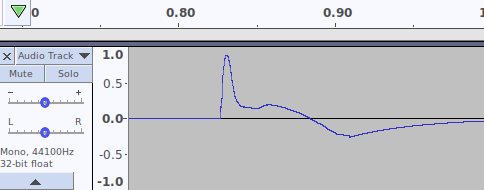}}
    \subfloat[Depiction of a saturated fsr peak \label{fig:peak-saturated}]{
        \includegraphics[width=0.5\textwidth]{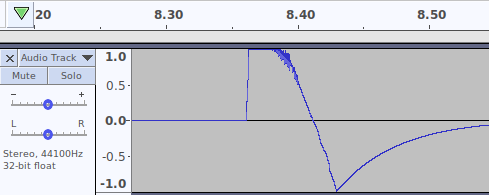}}
    \hfill
    \subfloat[Depiction of an inverted fsr peak \label{fig:peak-inverted}]{
        \includegraphics[width=0.5\textwidth]{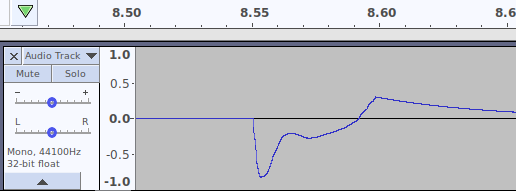}}
    \fnote{To test the
    setup, a recording of the computer's output and inputs on the device must
    be performed. The input device (in case of an external sound card) must be
    selected (a.a), recording should be started and stopped while the input
    device is operated (a.b), and then the recording should be played and a
    mixture of the sound being played and the tapping on the input device
    should be heard (a.c). A peak is expected to look as presented in fig
    \subref{fig:recorded-peak}. Fig \subref{fig:peak-saturated} presents the
    case where the peak saturated the recording, resulting in a flat top. Fig
    \subref{fig:peak-inverted} presents an inverted peak, where the first peak
    is negative.}
    \label{fig:audacity}
\end{figure}

The connections can easily be tested by playing an audio from the computer and
recording the audio while tapping on the input device. In figure
\ref{fig:audacity} we present part of the interface of an open-source sound
recording and editing software \citep{website:audacity}. By recording audio
from the recording device with the loopback connection, an audio track such as
the one in figure \ref{fig:recording_mix} should be produced. The track should
contain the stimulus signal on one channel and spikes corresponding the
tapping on the other one. 

This setup and software can also be used to inspect the
recording of the FSR signal to calibrate the FSR input device. Peaks are
expected too look as shown in figure \ref{fig:recorded-peak}. In case the FSR
signal saturates the audio card, the peak will contain a flat top as shown in
figure \ref{fig:peak-saturated}. This can be solved by adjusting the
potentiometer which regulates the maximum height of the peak. Another issue
might come from the sound card inverting the signal as in figure
\ref{fig:peak-inverted}. This can be solved by inverting the audio recording on
the recorded channel. A option to manage this is provided in the tool suit
described in \textbf{\nameref{sec:usage}}.

A website where further detail and tutorials are provided is detailed in the
\emph{\nameref{sec:open_practices}} section.

\section{Tool Suit and Usage Workflow}\label{sec:usage}
To make the use of the proposed setup as convenient as possible, in this work
we present a tool suit to produce the stimuli, record responses and analyze the
recordings. The tools are python programs open-sourced under an MIT licence.
The tool suit was developed using the UNIX Philosophy \citep{raymond2003art},
so each program is independent and dedicated to solving one issue. 

Next we outline the workflow considered and what are the tools we provide to
address each stage. An expected experiment design workflow would have the
following stages:

\begin{itemize}
\item \textbf{Define the stimuli}. Stimuli can be any arbitrary set of audios.
Many sensorimotor synchronization experiments use stimuli conformed by
discrete sound onsets at designated times. We provide an utility
(\texttt{beats2audio}) to transform a text file with onset times into an audio 
that produces a sound on each onset time.
\item \textbf{Expose participants to the stimuli and collect responses}. Given
a stimuli set, participants should hear each audio and produce responses
by tapping on the response device. The utility provided
(\texttt{runAudioExperiment}) receives a configuration file declaring the
audio stimuli set and presents each one while recording the response. Responses
for each trial are save as an audio file (fig \ref{fig:recording_mix}) on a
designated output folder.
\item \textbf{Extract tap times from the recordings}. Using the original
stimulus and the response recording, tap times are extracted relative to the
beginning of the stimulus. A different tool is provided for this purpose
(\texttt{rec2taps}).
\end{itemize}

In case the stimuli to be used is an audio with simple identical onsets on
designated times, the utility \texttt{beats2audio} receives a text file (figure
\ref{txt:beats}) with a list of onset times in milliseconds and outputs an audio
file (figure \ref{fig:recording_output}). A click sound is produced on each 
onset time. 

\begin{figure}
    \caption{\\ \textit{Example text file indicating onset times in milliseconds, one
    per line}}\label{txt:beats}
\centering
\begin{verbatim}
546
1638
1911
3003
3549
4914
\end{verbatim}
\end{figure}

\begin{figure}
\caption{\\ \textit{Depiction of a Trial.}}\label{fig:experiment_run}
\centering
\includegraphics[width=0.8\textwidth]{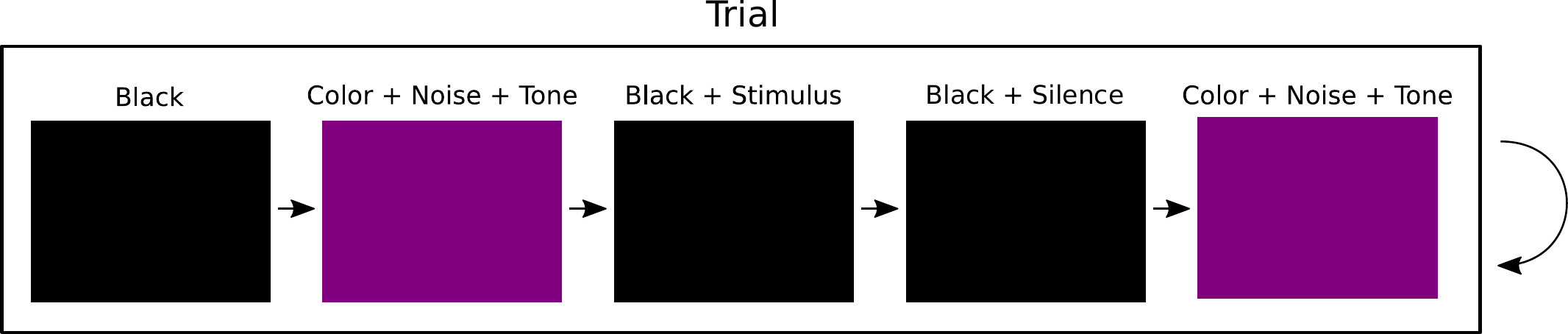}
\end{figure}

To run the tapping experiments, i.e.: producing the stimuli and recording the
loopback and input, we provide a simple utility named
\texttt{runAudioExperiment}. The utility requires three arguments: the path to
a configuration file, the path to a trial file and the path to an output
directory where recordings are to be stored.
The experiment execution follows the steps depicted in figure
\ref{fig:experiment_run}. Trials are run in a succession, each trial consisting
of five steps. First, a black screen is presented for an specified duration.
Secondly, screen turns to a (possibly) different color and white noise combined
with a tone is played. This option is intended to help remove rhythmic biases
between trials. Third, the screen goes black again and the trial stimulus is
played as recording is enabled. Fourth, the screen stays black in silence.
Recording continues in this stage. Finally, another optional colored noise
screen is produced. Following this screen, the next trial, starting with the
black screen, begins. 

The configuration file provided as the first argument of the utility specifies
the parameters of the execution of the steps mentioned above (figure
\ref{fig:run_config}). The trial file is a plain text file containing the
stimuli set, one per line as paths to audio files. The output directory defines
where the experiment outputs are to be saved. The utility produces as an output
one recording per trial, as obtained from the sound device, and a \texttt{csv}
file containing a table describing the details of the experiment execution
(table \ref{table:experiment_output}). The name of the output folder can be
used to identify experiment runs either by date, run id number or participant's
initials.

The last utility, \texttt{rec2taps}, extracts tap times from the audio
recordings produced during the experiment. To do so it requires two arguments,
a trial recording audio file and its original stimulus file. The utility uses
the original stimulus audio to find its starting point ($s^c$) in the
recording. It does so by looking for the maximum cross-correlation between the
stimulus and the loopback channel of the recording. Then, using the channel
where the input signal is recorded, the utility finds peaks in the signal and
extracts tap times as the location of the maximum of each peak. Given that the
beginning of the original stimulus can be found in the trial's recording, the
playback delay can be subtracted from tap times, obtaining tap times relative
to the beginning of the stimulus. Tap times are printed out in milliseconds,
one per line.

Detailed instructions on how to use the utilities described 
are provided in each code's
project readme files. Links to code projects are listed in the
\emph{\nameref{sec:open_practices}} section.
The next subsection informs in more detail how tap times
are extracted from the recording and analyzes its accuracy.

\begin{figure}
\caption{\\ \textit{Configuration file example. }}\label{fig:run_config}
\centering
\begin{verbatim}
black_duration: 600         # Duration of black screen (in ms)
c1_duration: 3000           # Duration of first noise screen (in ms)
c1_color: "#afd444"         # Color of first noise screen
c2_duration: 000            # Duration of second noise screen (in ms)
c2_color: "#afd444"         # Color of second noise screen
randomize: false            # Whether trial order should be randomized
sound_device: "default"     # String or int identifying the sound deviced used
silence_duration: 1500      # Duration of silence after stimuli playback
c_volume: 1.0               # Volume of cleaning sound
\end{verbatim}
\end{figure}

\begin{table}
\caption{\\ \textit{Example of an Output Table from an Experiment Run}}\label{table:experiment_output}
\centering
    {\scriptsize
\begin{tabular}{l | l | l | l | l | l | l}
index & stimulus\_path & recording\_path & black\_duration & c1\_duration &
    c2\_duration & silence\_duration \\
\hline
    0 & s1.wav & s1.rec.wav & 600 & 300 & 0 & 1000 \\
\end{tabular}
    }
\end{table}

\subsection{Signal Analysis}\label{sec:signal_analysis}

The goal of the setup is to be able to record a participant's tap times with
high precision. In the presented setup, the signal from the input device is a
function over time of the activation of the device, recorded as an audio
signal. From this signal, we intend to extract individual time points that
represent each actuation of the input device. In electronic input devices as
the one presented here (the FSR), the activation signal is not a simple on-off
function, but a curve describing the change of pressure on the device
over-time. To obtain individual tap times for each actuation, the signal must
be processed to recognize each activation and then select a point in time
within the curve that is representative of the time of actuation. 

This process raises two aspects subject to analysis. First, whether the
processing misses any individual activation or detects spurious activations.
Secondly, how representative the selected time point is within the activation
curve of the actuation process. 
Considering the functionality of the FSR, the signal peak is the 
moment where the highest pressure is applied to the input device.
We decided to select the maximum of the activation curve as a representation of
the tap time.
We will now focus our attention on the analysis of
the performance of the signal processing algorithm used in
\texttt{rec2taps} when applied to recordings performed with the
presented input device.

\begin{figure}
\caption{\\ \textit{Shape of an FSR Activation Peak.}}
\centering
\subfloat[Illustration of a single activation \label{fig:fsr_peak_s}]{
\includegraphics[width=0.5\textwidth]{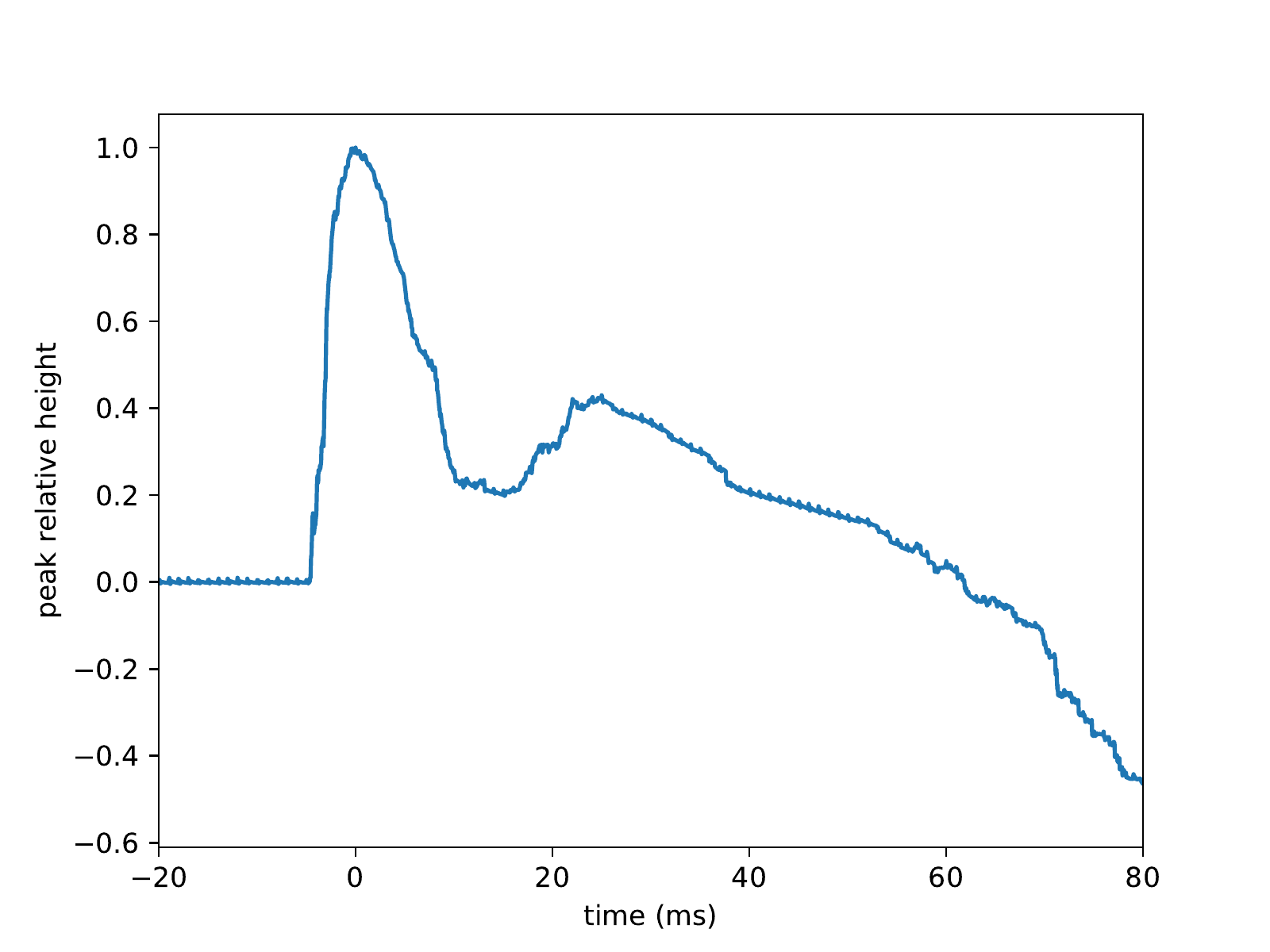}}
\subfloat[Illustration of a single activation after rectification \label{fig:fsr_peak_sr}]{
\includegraphics[width=0.5\textwidth]{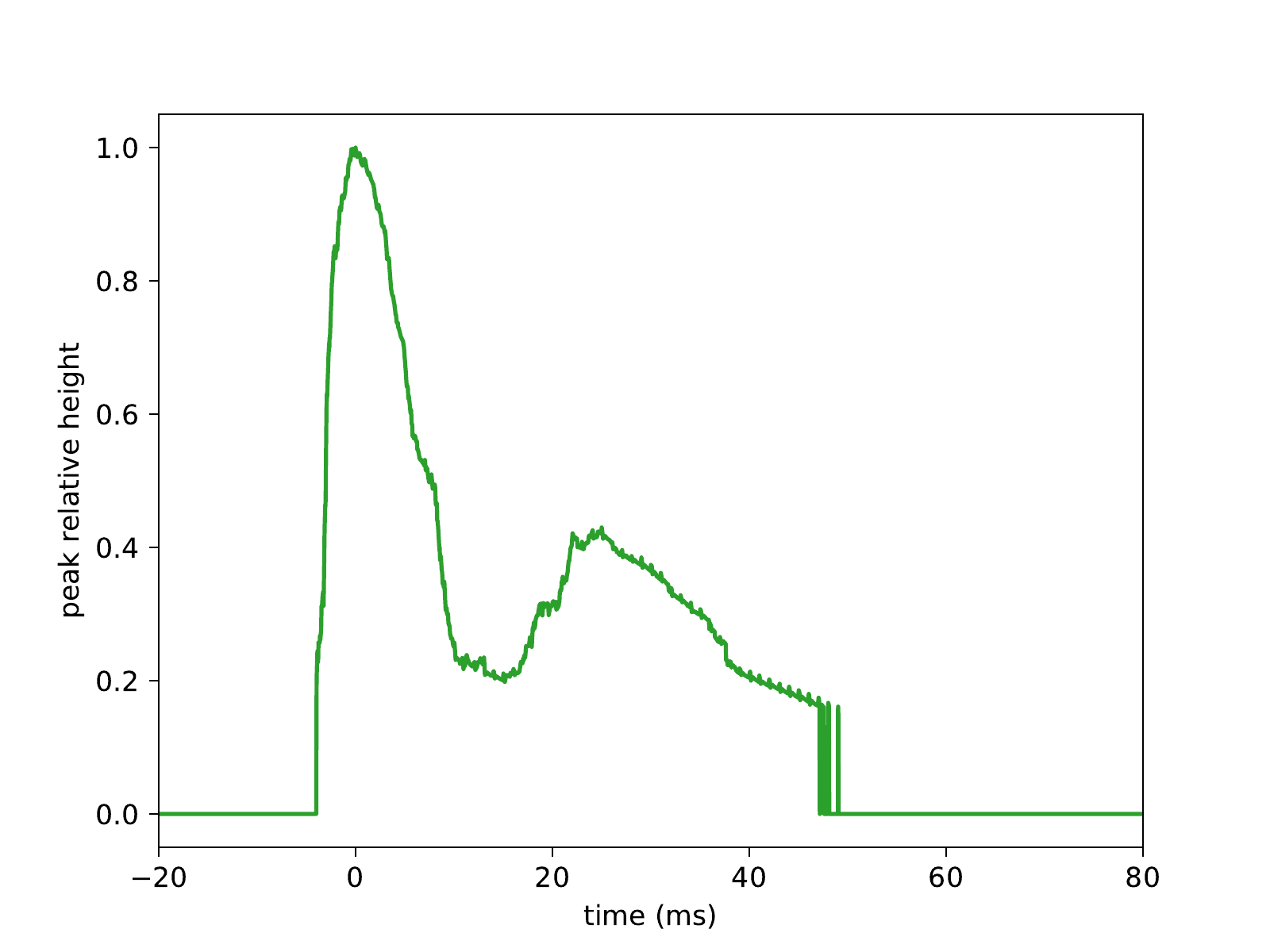}}
\hfill
\subfloat[Mean shape of an activation \label{fig:fsr_peak_d}]{
\includegraphics[width=0.5\textwidth]{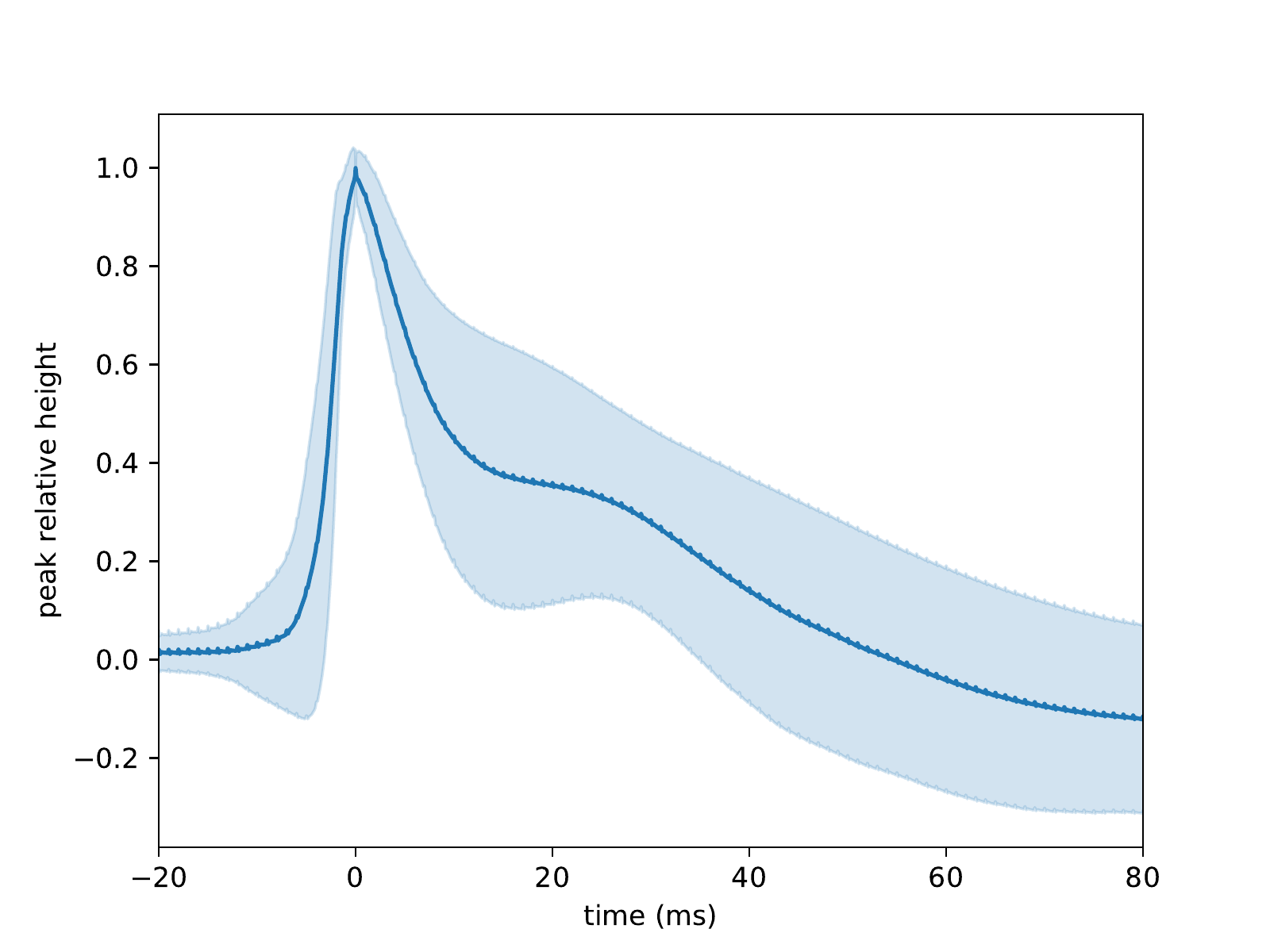}}
\subfloat[Mean shape of an activation after rectification \label{fig:fsr_peak_dr}]{
\includegraphics[width=0.5\textwidth]{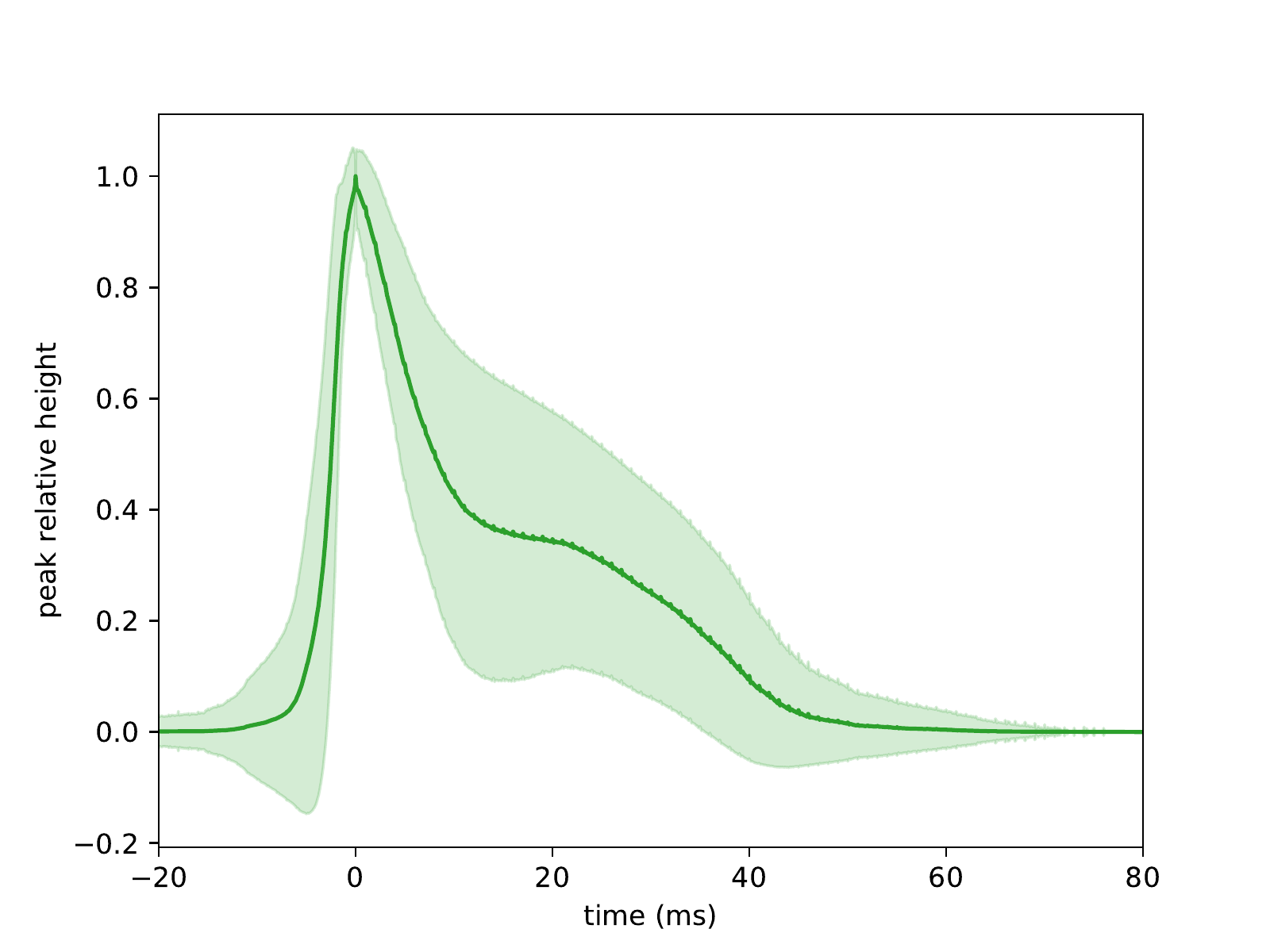}}
\hfill
\fnote{Main line shows mean relative
activation over-time. Time zero represents the maximum of the peak. Shading
represents standard deviation of the activation.}\label{fig:fsr_peak}
\end{figure}

The proposal of a Force Sensitive Resistor (FSR) as input device was due to the
clarity of the signal provided. The signal is very close to zero when it is not
being actuated and then raises rapidly proportionally to the pressure applied.
Figure \ref{fig:fsr_peak_s} shows the shape of one FSR activation when
aligned to the detected maximum and normalized to the peak's height. Figure
\ref{fig:fsr_peak_d} shows mean activation over time for 3000 peaks
from 68 recordings. The process to detect activations starts by
rectifying (setting to 0) the signal below a threshold defined as 1.5 times
the standard deviation of the signal amplitude (figure \ref{fig:fsr_peak_sr}).
Afterwards, peaks are found as points in the signal that are local maximums and
have a prominence of at least the mentioned threshold and are distanced between
each other at least 100 ms. The prominence of a peak measures the height
relative to the smallest valleys between a possible peak and any closest
greater peak.  Our utility uses the function \texttt{find\_peaks} from the
\texttt{scipy.signal} package (version 1.2.0) \citep{2020SciPy-NMeth}. 

To inspect the recall and over-sensitivity of the algorithm used, we inspected
its performance over a set of tapping recordings from a beat tapping experiment
using the proposed setup \citep{miguel2019tapping}. The experiment required
participants to listen to non-isochronous rhythmic passages performed by
identical click sounds and tap to a self-selected beat. Participants were free
to decide on the beat and were allowed to change the beat mid-rhythm or even
pause tapping. The data set comprises 518 recordings from
21 participants. The evaluation of the peak picking process
consisted on visually inspecting the recording signal with the detected tap
times overlapped and annotating for each recording the number of missing and
spurious activations. Inspection was performed by one of the authors. On
491 (105\%) of the audios no missing or
spurious activations were seen and only on 5
(\%) more than 2 missing or spurious activations were
reported. The experiment contained rhythms of varying complexity, some of which
had a very ambiguous beat. As a consequence, some participants produced taps of
varying strength, including some weak onsets. Whether these activations are to
be considered may depend on the nature of the experiment and can require making
the peak picking process more sensitive. The provided utility,
\texttt{rec2taps}, allows configuring the mentioned detection threshold as a
paramater. It also allows producing plots of the FSR signal and detected peaks
to calibrate the parameter. In the supplementary material we present plots as
the one produced by the utility as examples of tap detections for both general
cases and for cases with taps of varying strength to illustrate the workings of
the peak picking procedure.

\begin{figure}
\centering
\caption{\\ \textit{Distribution of Tap Strength for Delays of Filtered Signal
Maximum}}
\label{fig:fsr_asynchrony}
\includegraphics[width=0.8\textwidth]{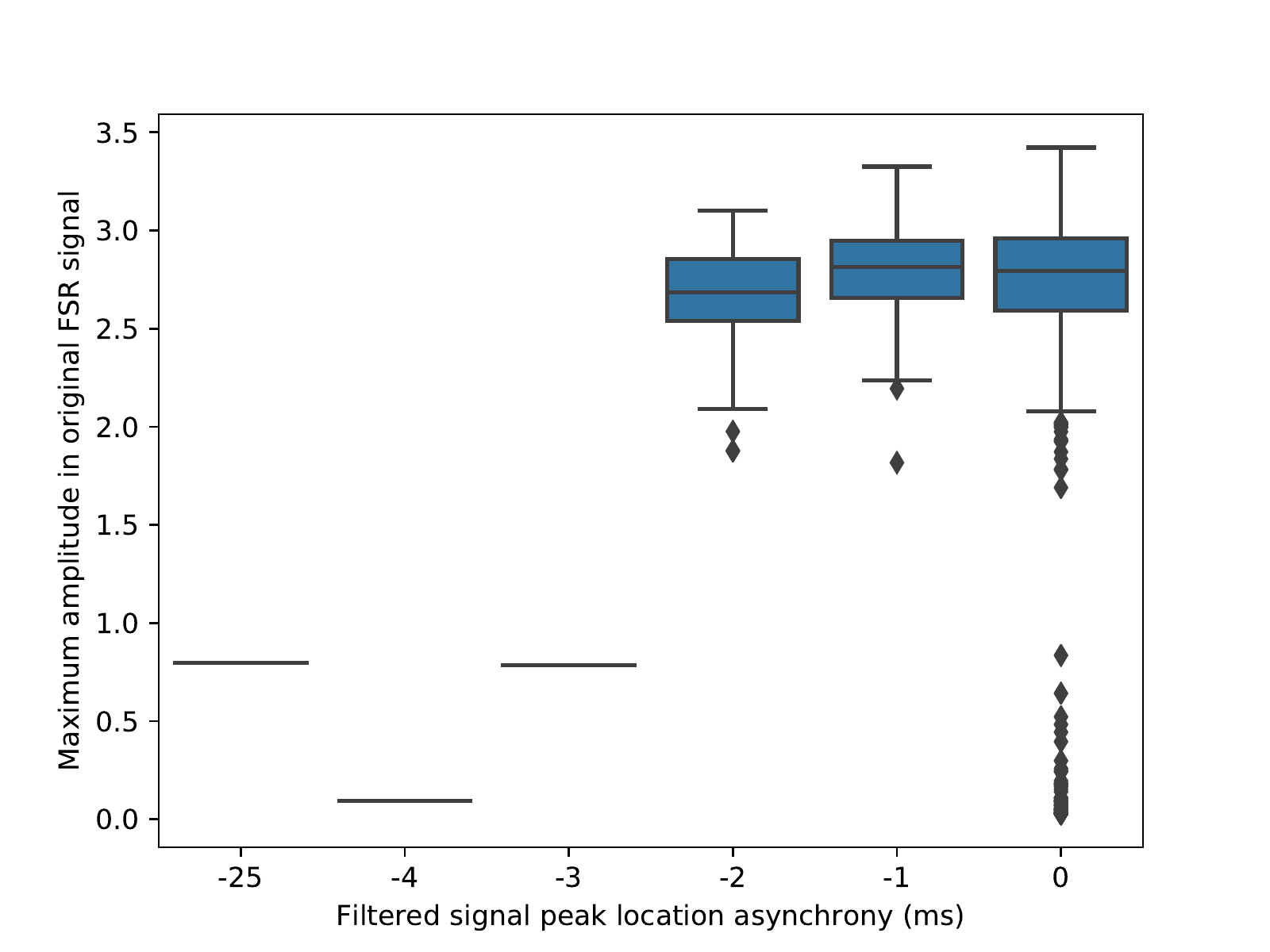}
\end{figure}

A final caveat of the signal processing regards the usage of a sound card as
the recording device. Sound cards generally high-pass filter the signal at
about 5Hz. As a consequence, the shape of the input signal may be modified,
specially if the tapping action was too soft. 
We examined the effect of a
high-pass filter on a FSR signal recorded with a digital signal acquisition device at 1000 Hz. 
To do so, we
resampled the original 1000 Hz signal to 48000 Hz using a linear interpolation,
applied a 5 Hz high-pass filter and recalculated the location in time of the
peak of the modified signal. We looked into 1508 tap activation profiles
from a synchronization experiment. In 67.57\% of the cases, the maximum of
the filtered signal remained in the same millisecond position. In 26.72\%
and 5.5\% of the cases, the maximum of the filtered signal was one or
two milliseconds ahead of the original signal, respectively. In 0.2\%, the
shift was greater than two milliseconds, up to -25 ms. We hypothesized
the negative lag of the maximum of the filtered signal to be related with a
soft tapping action. We inspected this hypothesis by looking at the maximum
amplitude of the FSR signal with respect to the peak's lag. Effectively, lags
greater than 2 ms were seen only in taps three times softer than average.
Figure \ref{fig:fsr_asynchrony} presents the distribution of the amplitude of
the peaks for each lag found.

\section{Discussion}\label{sec:discussion}
The current work presents an experimental setup intended to collect timed
responses with high-precision (less than -3ms delay) synchronized to onset
times in auditory stimuli. Another main characteristic of the proposed setup
is its inexpensiveness and simplicity. The introduction presents the general
schematic of experimental trials with auditory stimuli where the quantity of
interest is elapsed time from the moment a stimuli is heard until a response is
provided. In the \textbf{\nameref{sec:problem_description}} section we
explained why this quantity cannot be measured in an standard experimental
setup using a computer.  We also review previous approaches to the issue. In
the \textbf{\nameref{sec:setup}} section we describe our proposed setup and
provide detailed instructions for assembly. Finally, in
\textbf{\nameref{sec:usage}} we present an open-source software tool suit to
run experiments using the presented setup. 

Being able to collect response times to auditory stimuli with precision cannot
be easily done using a standard computer with default input devices (keyboard
or mouse). This is due to latencies introduced between the input device and the
experiment software or between the experiment software and the output device.
Approaches to this issue are either using specialized hardware for input and
output, running the experiment using a programmable micro-controller or
recording auditory output and responses in a separate recording device. Our
setup uses the last method.

Although this approach has been presented before in \cite{elliott2009mattap}, we
here present a less expensive alternative by using an inexpensive recording
device.  Moreover, we present assembly instructions for an inexpensive input
device that allows high precision recording. This setup easily allows using any
audio as stimulus. In addition, we provide a open-source tool suit for using
the setup that does not require proprietary software.

An evaluation of the performance of the software's capability to detect
participant's responses is described at the end of the previous section. The
evaluation showed no spurious or missing activation detections in 94\% of the
analyzed recordings and under 1\% presented more than two. Miss-detection of
activations was seen to relate to situations where the tapping action varied in
strength throughout the experiment trial.

A main limitation of this setup is that it cannot respond to participants
responses, either by changing the course of the experiment or providing
feedback. In that situation, most inexpensive approach is given in
\cite{schultz2019roles}. In case of access to precise input equipment, more
direct setups can be accomplished \citep{finney2001ftap, bridges2020timing}.
Another caveat of the setup presented here is a possible shift in
response time in case of soft tapping (see subsection
\nameref{sec:signal_analysis}). Finally, assembly of the input device requires
a minimum knowledge of electronics.

In summary, we provide an inexpensive setup for recording responses to auditory
stimuli with millisecond precision together with a software tool suit for using
the setup. The main focus is on getting high-precision response times relative
to the auditory stimulus with minimal calibration. 
 
\section{Acknowledgments}
This work was carried out by the corresponding author under a PhD Scholarship
provided by CONICET. No conflict of interests are declared.

\section{Open Practices Statements}\label{sec:open_practices}
The software tool suit is available with an MIT license via github repositories.
Data used to evaluate the precision of the tap extraction algorithm is
available upon request. None of the experiments were pre-registered.

Links to open source content:
\begin{itemize}
    \item Detailed instructions, tutorials and discussions on setup assembly:
        \url{https://github.com/m2march/tapping_setup}
    \item Code for \texttt{beats2audio} utility:
        \url{https://github.com/m2march/beats2audio}
    \item Code for \texttt{runAudioExperiment} utility:
        \url{https://github.com/m2march/runAudioExperiment}
    \item Code for \texttt{rec2taps} utility:
        \url{https://github.com/m2march/rec2taps}
\end{itemize}

\newpage
\section{Supplementary Material}

\begin{figure}[h]
    \caption{\\ \textit{Sample Depiction of Average Tap Detection Scenarios.}}
\includegraphics[width=\textwidth]{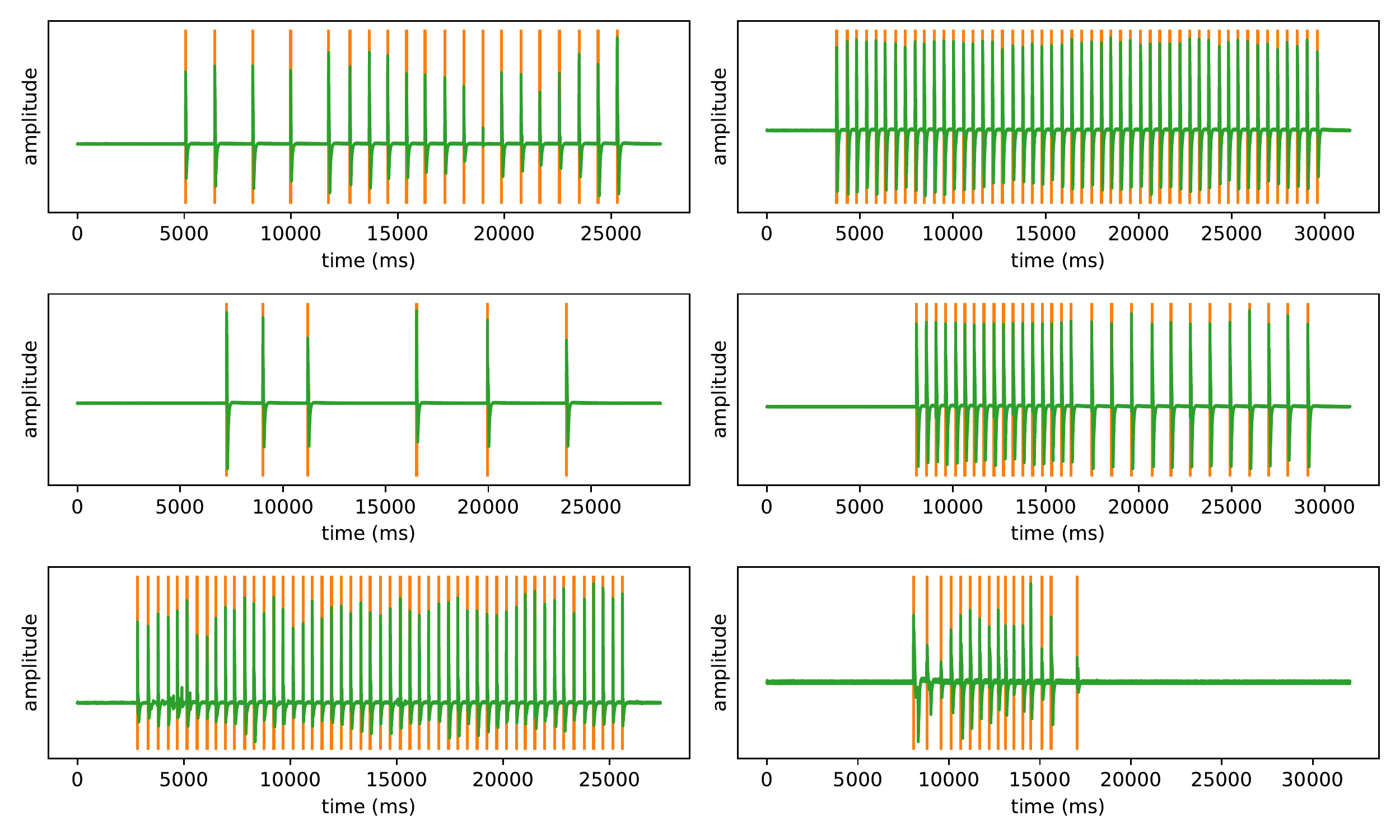}
\fnote{Signal (green) is the activation of the input device (FSR) during one
trial. Vertical lines (orange) are detected taps.}
\end{figure}

\newpage

\begin{figure}[h]
    \caption{\\ \textit{Depiction of Worst Tap Detection Scenarios}}
\includegraphics[width=\textwidth]{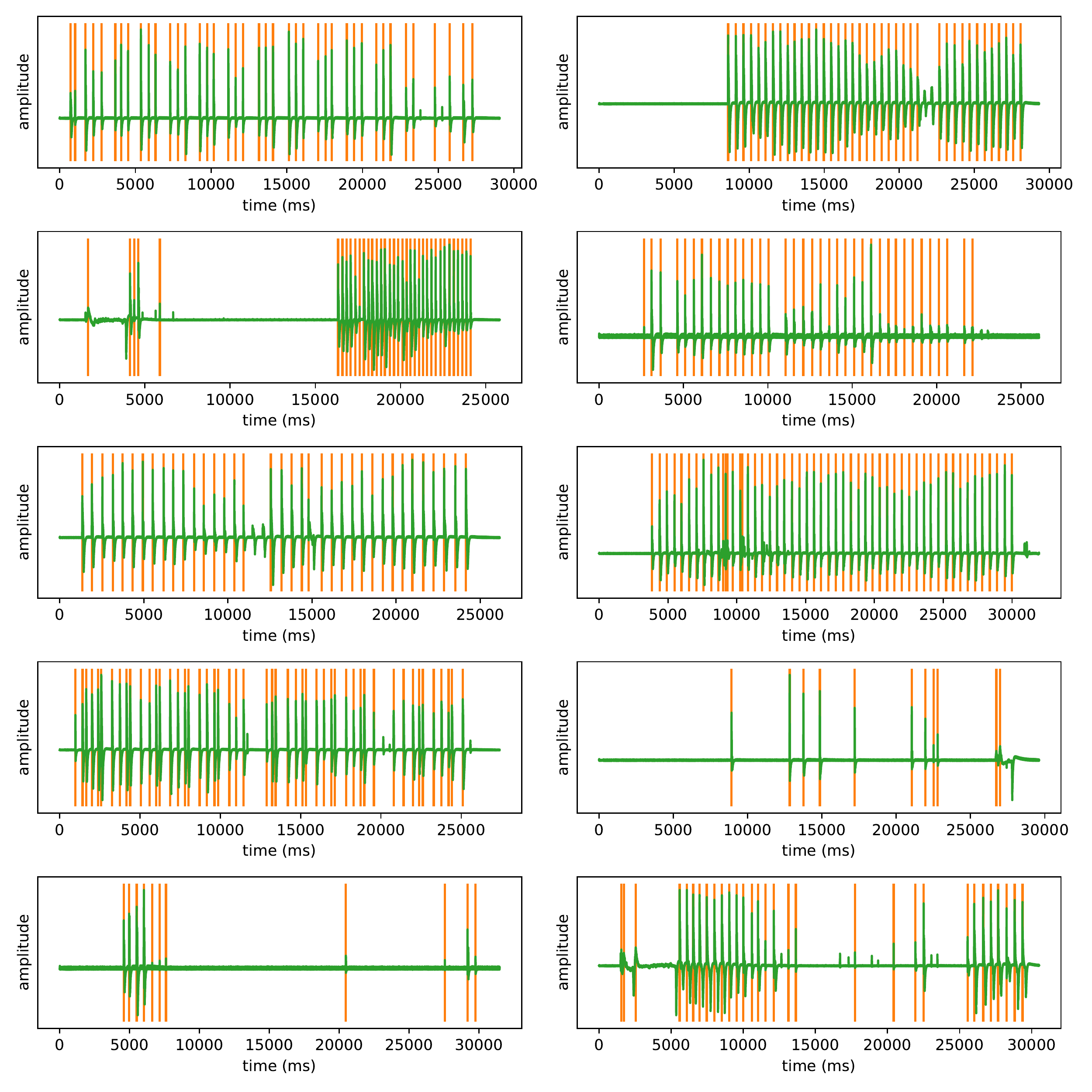}
\fnote{Worst tap scenarios are considered as higher ratio of
missing or spurious taps over detected taps. In most cases, missing taps are
soft taps in comparison to the rest. Signal (green) is the activation of the
input device (FSR) during one trial. Vertical lines (orange) are detected
taps.} 
\end{figure}

\newpage

\end{document}